\documentclass[12pt]{article}
\usepackage[mathscr]{eucal}
\usepackage{bm}
\usepackage{bbold}
\usepackage{caption}
\usepackage{multirow}
\usepackage{amsmath}
\usepackage{textgreek}
\usepackage[greek,english]{babel}
\usepackage{xcolor}
\usepackage{graphicx}
\usepackage{empheq}
\usepackage{axodraw2}
\usepackage{amsfonts}
\usepackage{amssymb}
\usepackage{dsfont}
\usepackage{ytableau}
\usepackage{overpic}
\usepackage{hyperref} 
\usepackage{makeidx}
\usepackage{empheq}

\newcommand{\be}{\begin{eqnarray}}
\newcommand{\ee}{\end{eqnarray}}
\newcommand{\bc}{\begin{center}}
\newcommand{\ec}{\end{center}}
\newcommand{\nn}{\nonumber \\}
\newcommand{\lb}{\label}
\newcommand{\p}[1]{(\ref{#1})}
\newcommand{\vecg}[1]{\mbox{\boldmath $#1$}}
\renewcommand{\u}{\underline}

\begin{document}

\begin{titlepage}

\vspace*{0.2cm}

\begin{center}

{\LARGE\bf  Symmetries of massless QCD}

\vspace{2cm}

{\Large Andrei Smilga} \\

\vspace{0.5cm}

{\it SUBATECH, Universit\'e de
Nantes,  4 rue Alfred Kastler, BP 20722, Nantes  44307, France. }

\end{center}
\vspace{0.2cm} \vskip 0.6truecm \nopagebreak

   \begin{abstract}
We present a pedagogical review of certain exact theoretical results concerning the physics of 
an imaginary world where one quark or more are deprived of their masses. 
   \end{abstract}

\end{titlepage}

\tableofcontents

\setcounter{footnote}{0}

\setcounter{equation}0
\newcommand{\vecind}[1]{\mbox{\scriptsize \boldmath $#1$}}
\newcommand{\ga}{\lower.7ex\hbox{$
\;\stackrel{\textstyle>}{\sim}\;$}}
\newcommand{\la}{\lower.7ex\hbox{$
\;\stackrel{\textstyle<}{\sim}\;$}}
\newcommand{\beq}{\begin{equation}}
\newcommand{\eeq}{\end{equation}}
\newcommand{\pd}{\partial}
\newcommand{\bebox}{\begin{empheq}}
\newcommand{\ebox}{\end{empheq}}
\renewcommand{\u}{\underline}
\renewcommand{\vec}[1]{{\bf #1}}
\newcommand{\red}{\color{red}}
\newcommand{\blue}{\color{blue}}

\section{Introduction} 

Quantum chromodynamics is the theory of strong interactions. Its Lagrangian includes fundamental quark and gluon fields. It reads
\be
\label{LQCD}
 {\cal L}_{QCD}  = - {1\over 2g^2} {\rm Tr} \{{\hat F}_{\mu\nu} {\hat F}^{\mu\nu} \}
 \ +\  \sum_{f=1}^6 \bar\psi_f (i{\cal D}\!\!\!\!/ - m_f) \psi_f \ +\ \frac \theta{32\pi^2} \varepsilon^{\mu\nu\alpha\beta} 
 {\rm Tr} \{{\hat F}_{\mu\nu} {\hat F}_{\alpha\beta} \}
   , \ee
where 
\be
  \label{Fnab}
\hat F_{\mu\nu}   = \partial_\mu \hat{A}_\nu - \partial_\nu \hat{A}_\mu
  - i[\hat{A}_\mu, \hat{A}_\nu] = i[{\cal D}_\mu, {\cal D}_\nu]
 \ee
with ${\cal D}_\mu = \pd_\mu - i \hat A_\mu$ are gluon field densities belonging to the adjoint representation of the color $SU(3)$ gauge group; $\psi_f$ are the quark fields lying in the fundamental representation of $SU(3)$; $g$ is the strong coupling constant, the experimental restriction on the parameter $\theta$  in the last term in \p{LQCD}  breaking the $P$ and $T$ symmetries 
is\footnote{It follows from the restrictions on the observed value of the neutron electric dipole moment.} $|\theta| \la 10^{-10}$.

Now, $m_f$ are the quark masses. Three quarks are heavy: their pole masses (the positions of the poles in the perturbative quark propagators) are $m_c \approx 1.35$ GeV, $m_b \approx 4.8$ GeV and $m_t \approx 170$ GeV. Three other quarks are relatively light:
 $m_u \approx 3$ MeV, $m_d \approx 5$ MeV and $m_s \approx 100$ MeV. $m_u, m_d$ and $m_s$ are essentially less than the characteristic hadron scale $\mu_{\rm hadr} \sim 500$ MeV.
Bearing the latter fact  in mind, it is theoretically interesting to inquire what would happen if these masses would be absent. What would be the physics of such imaginary world? 

This question is discussed in the present review. It covers the material well-known to experts in a concise and hopefully pedagogical way. It was written on the basis of Lectures 12,13,14 of the book \cite{Lectures}, but some extra discussions have been added.

We will assume in the following that $\theta = 0$. The physics of an imaginary world with nonzero $\theta$ is also rather nontrivial, and we dare address an interested reader to Lectures 15 and 16 of Ref. \cite{Lectures} or to the original papers \cite{Leutwyler,tetpi}. 

Obvious symmetries of QCD, massless or not, are the global Poincare symmetry and the local gauge symmetry:
\be
   \label{gaugefin}
   &&\hat{A}_\mu \ \to \ \Omega(x) \hat{A}_\mu \Omega^\dagger(x) - i
\left[ \partial_\mu
   \Omega(x) \right] \Omega^\dagger (x), \nn
&& \psi_f(x) \ \to \  \Omega(x) \psi_f(x) 
 \ee
with $\Omega(x) \in SU(3)$.

However, the gauge symmetry is actually
{\it not} a symmetry in the same sense as  rotational or Lorentz symmetry are. 
Namely, in the case of gauge symmetry, we  are not allowed
to consider  states which are not invariant under symmetry transformations;
it does not act on the Hilbert space
of  physical states, which are all  gauge singlets annihilated by the generators of gauge transformations.
Gauge symmetry exhibits itself only in the Lagrangian formulation,
involving some extra unphysical variables which can in 
principle be disposed of. One can say that  gauge symmetry is not a symmetry, but rather
a convenient way to describe constrained systems. 

We will concentrate here on  global symmetries specific for {\it massless} QCD. These are {\it conformal symmetry} and {chiral symmetries}.

\setcounter{equation}0

\section{Conformal symmetry and its breaking.}
\setcounter{equation}0

The  bare Lagrangian (\ref{LQCD}) of QCD with strictly massless quarks involve no scale.
 That means that the classical action is invariant
under the transformations
 \be
\label{scaleA}
\left\{ \begin{array}{l}
x^\mu \to \lambda x^\mu \\
\hat A_\mu \to \lambda^{-1} \hat A_\mu \\
\psi_f \to \lambda^{-3/2} \psi_f
\end{array} \right. \ .
\ee
Here we took into account the fact that the gauge field $A_\mu$ has the 
canonical dimension of mass and the canonical dimension of the quark field is
$[\psi] = m^{3/2}$.

For  pure gauge theory,  a somewhat stronger form of the scaling symmetry 
(\ref{scaleA}) holds. Consider the action of  Yang--Mills theory
on a curved four-dimensional manifold
  \be
 \label{YMcurv}
S_{\rm curved}^{\rm YM} \ =\ - \frac 1{2g_0^2} {\rm Tr} \int d^4x \sqrt{-
{\rm det} \|g\|}\  
\hat F_{\mu\nu}
\hat F_{\alpha\beta} g^{\alpha\mu} g^{\beta\nu},
  \ee
where $g_0^2$ is the coupling constant, not to be confused with 
  the metric tensor $g_{\mu\nu}(x)$.\footnote{We assume that the reader is familiar with the basic notions of
Riemannian geometry. Eq.~(\ref{YMcurv}) is an obvious non-Abelian generalization
for the action of an electromagnetic field on a curved background (see e.g. 
Ref. \cite{LLFT}).}
Note now that the action (\ref{YMcurv}) is invariant under the local conformal
 transformations
  \be
\label{locCT}
\left\{
\begin{array}{l}
g_{\mu\nu}(x) \ \to \ \lambda^{-2}(x) g_{\mu\nu}(x) \\
g^{\mu\nu}(x) \ \to \ \lambda^{2}(x) g^{\mu\nu}(x) 
\end{array} 
\right. \ ,
   \ee
while the fields and coordinates are not transformed. For
a  flat metric $g_{\mu\nu}(x) = \eta_{\mu\nu}$  
and $\lambda(x) = \lambda$, the transformation (\ref{locCT}) amounts to a homogeneous scale dilatation
  and is equivalent to (\ref{scaleA}).\footnote{Note in parentheses that the
local conformal invariance is not specific to gauge theories. Some other Weyl-invariant theories exist. Einstein gravity is not Weyl-invariant, but a variant of quadratic gravity, the {\it Weyl gravity} is.  Its action reads
\be
S^{\rm Weyl}  \ \propto \ \int d^4x \, C^2_{\mu\nu\rho\sigma} \sqrt{-{\rm det} \|g\|},
\ee 
where the {\it Weyl tensor}  is
$$
C_{\mu\nu\rho\sigma} \ =\ R_{\mu\nu\rho\sigma} - \frac 12 (g_{\mu\rho} R_{\nu\sigma}  +g_{\nu\sigma} R_{\mu\rho}  - g_{\mu\sigma} R_{\nu\rho} - g_{\nu\rho} R_{\mu\sigma} )
+ \frac R6(g_{\mu\rho} g_{\nu\sigma} - g_{\nu\rho} g_{\mu\sigma})
$$
($R_{\mu\nu}$ is the Ricci tensor and $R = R^\mu_\mu$ is the scalar curvature).

Also the theory of scalar field coupled to gravity in the following way,
\be
S   \ \propto \  \int d^4x \, \left(\frac 1{12} R\phi^2 + \frac 12 g^{\mu\nu} \pd_\mu \phi \pd_\nu \phi \right)\sqrt{-{\rm det} \|g\|},
\ee
enjoys Weyl invariance.
}

As far as  Yang--Mills theory is concerned, the symmetry (\ref{locCT}) has 
an important consequence:\footnote{Cf. Eq. (94.5) in the book \cite{LLFT}.}
  \be
\label{traceYM}
0 \ = \ \frac {\delta S_{\rm curved}^{\rm YM}}{\delta \lambda(x)} \ \propto \ 
 g_{\mu\nu}(x) \frac {\delta S_{\rm curved}^{\rm YM}}{\delta g_{\mu\nu}(x)}
\ =\ - \frac 12 \Theta_\mu^\mu(x)\sqrt{-{\rm det}\|g\|},
\ee
where $\Theta^{\mu\nu}(x)$ is (the symmetric version of) 
the energy--momentum tensor:

 \be 
\label{Tmunu}
\Theta^{\mu\nu} \ =\ \frac 2{g_0^2} {\rm Tr} \left[- \hat F^{\mu \rho} 
\hat F^{ \nu}_{\ \rho}\  +\ 
\frac {g^{\mu\nu}} 4 \hat F_{\rho\sigma} \hat F^{\rho\sigma} \right] .
 \ee
To derive \p{Tmunu}, we used the language of Riemannian geometry, but it holds also in
flat space with $g_{\mu\nu}(x) = \eta_{\mu\nu}$ we are 
primarily interested in. Then the divergence $\pd_\mu \Theta^{\mu\nu}$ vanishes --- this is the energy and momentum conservation.

Note that the expression for $\Theta^{\mu\nu}$ derived in the framework of flat canonical formalism  by  Noether's method
  does not coincide in  form with \p{Tmunu}. For example, for pure photodynamics with ${\cal L} = - F_{\rho\sigma}F^{\rho\sigma}/4$, one obtains:
 \be 
\label{Tcan}
\left(\Theta^{\mu\nu}\right)^{\rm can} =
\frac {\delta {\cal L}}{\delta (\partial_\mu A_\rho)} \partial^\nu A_\rho
- \eta^{\mu\nu} {\cal L} = - F^{ \mu \rho}
 \partial^\nu A_\rho\, +\,
\frac {\eta^{\mu\nu}} 4 F_{\rho\sigma} F^{\rho\sigma}.
 \ee
In contrast to Eq.~(\ref{Tmunu}), this expression is not symmetric under interchange $\mu \leftrightarrow \nu$ 
({\it \`a propos}, it is not gauge invariant either). We have:
$$  (\Theta^{\mu\nu})^{\rm can} \ =\ (\Theta^{\mu\nu})^{\rm sym} - F^{\mu\rho} \pd_\rho A^\nu =
(\Theta^{\mu\nu})^{\rm sym} - \pd_\rho (F^{\mu\rho}  A^\nu), 
$$
where we used the equation of motion $\pd_\rho F^{\mu\rho} = 0$ of pure photodynamics.
As $\partial_\mu\partial_\rho (F^{\mu\rho} A^\nu) =  0$ by (anti)symmetry, both
$(\Theta^{\mu\nu})^{\rm can}$ and $(\Theta^{\mu\nu})^{\rm sym}$ are conserved: the forms \p{Tmunu} and \p{Tcan} are equivalent. 

  The relation $\Theta^\mu_\mu = 0$ following from \p{traceYM}
can be interpreted as a local conservation law of the dilatation
 current
  \be
J_D^\mu \ =\ x_\nu \Theta^{\mu\nu}.
 \label{curdil}
\ee
 To understand better the meaning of \p{curdil}, 
let us derive the canonical expression for the dilatation
 current for a general theory with  
Lagrangian ${\cal L}(\phi_i, 
\partial_\mu \phi_i)$, depending on some set of bosonic fields $\phi_i(x)$ with
canonical dimension 1, and their derivatives. To this end, 
consider an infinitesimal scale
 transformation 
  \be
\label{scalex}
\left\{
\begin{array}{l}
  \delta \phi_i = \alpha \phi_i  \\
\delta x^\mu = - \alpha x^\mu 
\end{array} \right.
\ee 
($\alpha \ll 1$) and rewrite this  in a form where 
only the fields, but not the coordinates, are transformed. We have
 \be
\label{scaleinf}
\delta \phi_i &=& \alpha(x^\nu \partial_\nu \phi_i + \phi_i)  \nonumber\\
\stackrel{\alpha = {\rm const}}{\Longrightarrow} \ \ \ 
\delta(\partial_\mu \phi_i) &=& \alpha(x^\nu \partial_\nu \partial_\mu \phi_i +
2\partial_\mu \phi_i),
 \ee
 where the first terms in $\delta \phi_i(x)$,  $\delta [\partial_\mu \phi_i(x)]$
 compensate for the shift of argument
and the second terms reflect the  canonical dimension 1 of the fields $\phi_i$ 
and the canonical dimension 2 of the fields $\partial_\mu \phi_i$.
 The variation of the Lagrangian under the global transformation (\ref{scaleinf})
is
  \be 
\label{vardilL}
\delta {\cal L} \ = \  \alpha \left\{ x^\nu \partial_\nu {\cal L} 
+ \sum_i \left[ \frac{\delta {\cal L}}{\delta \phi_i} \phi_i +
2 \frac{\delta {\cal L}}{\delta (\partial_\mu \phi_i)} \partial_\mu \phi_i 
\right] \right\}.
 \ee
For a  Lagrangian of canonical dimension 4, the second term is just $4\alpha {\cal 
L}$  and the variation \p{vardilL} boils down to 
a total derivative $\delta {\cal L}\ =  \alpha \partial_\nu
(x^\nu {\cal L})  \ \stackrel {\rm def} = \alpha \pd_\nu f^\nu$. Assume now that the parameter $\alpha(x)
$ is not a constant and calculate the canonical Noether
current 
\be
J^\mu \ =\ \frac {\delta {\cal L}}{\delta (\pd_\mu \alpha)} - f^\mu .
 \ee
 The variation of $\pd_\mu \phi_i$ in Eq. \p{scaleinf} acquires  
an extra term:
$$ \delta(\partial_\mu \phi_i) = \alpha(x^\nu \partial_\nu \partial_\mu \phi_i +
2\partial_\mu \phi_i) + \pd_\mu \alpha (x^\nu \pd_\nu \phi_i + \phi_i).  
$$
Then
$$ \frac{\delta {\cal L}}{\delta (\pd_\mu \alpha)} \ =\  \sum_i \frac{\delta {\cal L}}{\delta (\pd_\mu \phi_i)} (\phi_i + x^\nu \pd_\nu \phi_i) 
$$
and

  \be
 \label{Dcan}
(J_D^\mu)^{\rm can} \ =\ \sum_i \phi_i
\frac{\delta {\cal L}}{\delta (\partial_\mu \phi_i)} \ +\ x_\nu 
(\Theta^{\mu\nu})^{\rm can},
 \ee
where
$$
(\Theta^{\mu\nu})^{\rm can} \ =\ 
 \sum_i \frac{ \delta {\cal L}}{\delta (\partial_\mu \phi_i)} \partial^\nu \phi_i
- \eta^{\mu\nu} {\cal L}
$$
is the canonical energy--momentum tensor.
 For  photodynamics the expression (\ref{Dcan}) differs from
$x_\nu (\Theta^{\mu\nu})^{\rm sym}$ by the term
  \be
\Delta J^\mu_D \ =\ (J_D^\mu)^{\rm can} - x_\nu (\Theta^{\mu\nu})^{\rm sym} \ =\ 
-F^{\mu\rho} A_\rho - x_\nu F^{\mu\rho} \partial_\rho A^\nu
\label{delD}. 
\ee
Taking into account the equations of motion $\partial_\rho F^{\mu\rho} =0$,
we obtain $\Delta J^\mu_D\ =\ -\partial_\rho(x_\nu F^{\mu\rho} A^\nu)$, i.e.\ 
the canonical dilatation
 current (\ref{Dcan}) coincides with the current
(\ref{curdil}) up to a term whose divergence is zero. Thus,
one can use the current (\ref{curdil}) instead of (\ref{Dcan}) in all 
 calculations. 

We presented a quite explicit demonstration of this fact for photodynamics, and a similar explicit demonstration for  non-Abelian theory (I did not find it in the literature) may be of  a certain methodic interest. An underlying reason for this equivalence is  the 
local conformal invariance of the action (\ref{YMcurv}) in a curved
background.

For a theory involving also massless quarks, an extra term 
 \be
\label{traceq}
\left(\Theta_\mu^\mu \right)^q \ =\
 -3i \sum_f \bar\psi_f /\!\!\!\!{\cal D} \psi_f
  \ee
appears in the trace.
 It is not zero off-shell, but
it is still zero for  fields satisfying the classical 
equations of motion 
$ /\!\!\!\!{\cal D} \psi_f = 0$. For massive quarks, scale invariance is lost 
and $\Theta_\mu^\mu  \ \sim\
  \sum_f m_f \bar\psi_f \psi_f \ \neq 0$.
 
\vspace{1mm}

Up to now, we were  only discussing the classical theory. What happens in the 
quantum case? One may ask whether  conformal symmetry is still present and, 
in particular, whether the trace of the quantum {\it operator} corresponding
to the classical expression (\ref{Tmunu}) for the energy--momentum tensor
still vanishes?

 First of all, note that the neglection of  terms which vanish due to the classical
equations of motion is quite justified. The counterpart of the latter in 
quantum theory are the Heisenberg operator
 equations of motion
 which tell us that
all the matrix elements of  operators like $ (i\,/\!\!\!\!{\cal D} - m) \psi $
between the physical states vanish. Still, 
the answer to the question above is {\it negative} due to the phenomenon of
 of  {\it dimensional transmutation}.

As is well known, the coupling constants in QED or in QCD  are not constant --- they {\it run} depending on a characteristic energy of a given process.
In QCD, they go down with the energy: this is what is called {\it asymptotic freedom}. In leading logarithmic approximation, the energy dependence of $\alpha_s(E) = g^2(E)/4\pi$  has the form 
\be
 \label{asfred}
 \alpha_s (E) \ =\ \frac {2\pi}{b_0 \ln \frac {E} {\Lambda_{\rm QCD}}}
 \ , \ee 
where 
\be
\lb{b0} b_0\  = \ \frac{11N_c - 2N_f}3
 \ee
 ($N_c$ being the number of colors and $N_f$  the number of light quark flavors) and $\Lambda_{\rm QCD} $ [rather than running  $g(E)$ or $\alpha_s(E)$] is the real fundamental constant of quantum chromodynamics. Experiment gives the value $\Lambda_{\rm QCD} \approx $ 200  MeV. 

According to \p{asfred}, $\alpha_s(\Lambda_{\rm QCD})$ become infinite, but this perturbative formula is not valid when $\alpha_s$ becomes large. 
One should rather say that at some   energy scale less than 1 GeV, the strong interaction becomes really strong so that perturbative calculations are not possible anymore and the physics cannot be described in terms of quarks and gluons. At low energies, the latter are hidden ({\it confined}) in the colorless hadron states, the lowest such states  having the mass of order of $\Lambda_{\rm QCD}$ multplied by a numerical factor. 
 Even though we cannot prove it theoretically and the corresponding Millennium Prize is not awarded yet, we see that in experiment.   
What is important for us in this review is the fact that conformal symmetry of pure Yang--Mills theory or of massless QCD is there at the classical level, but this symmetry is broken by quantum effects, is {\it anomalous}.

We derived the existence of fundamental intrinsic energy scale in QCD based on the perturbative formula \p{asfred}, which could be derived using operator approach. But a more clear understanding of this phenomenon may be achieved using path integral language. To attribute meaning to path integral symbol, one has to {\it regularize} it  replacing an infinite-dimensional integral  by a finite dimensional one. The most natural way to do so is to put the theory on a {\it lattice} of finite size\footnote{Technical details with be given in Sect. 6.} Such lattice has a scale  $a$ --- the distance between two adjacent nodes. The corresponding energy scale $a^{-1}$ has the meaning of {\it ultraviolet regulator} $\Lambda_{UV}$. It is an artificial unphysical scale introduced in theory by hand. But it is this ultraviolet regulator which together with the bare coupling constant $g_0$ entering the QCD Lagrangian \p{LQCD} determines quite physical and important scale $\Lambda_{\rm QCD}$. It follows from Eq. \p{asfred} that, in the leading logarithmic order,
\be
 \label{dimtran}
 \Lambda_{\rm QCD} \ = \  \Lambda_{UV} \exp\left\{ - \frac {8\pi^2}{b_0 g_0^2}
 \right\}.
 \ee
And this is what is called dimensional transmutation.

The relation \p{dimtran} takes into account only leading logarithms. An improved version of this relation that includes all perturbative corrections reads\footnote{This result follows from the solution of the functional {\it renormalization group} equations. This issue is elucidated in many textbooks (e.g. in Chapter 12 of the book \cite{Pes} or in Lecture 9 of the book \cite{Lectures}), where we are addressing the reader.}
\be
 \label{dimtranexact}
 \Lambda_{\rm QCD} \ = \  \Lambda_{UV} \exp\left\{  \int_{g_0^2}^\infty \frac {dt}{\beta(t)} \right\}.
 \ee
The {\it Gell-Mann-Low function} $\beta(t)$ is defined as the logarithmic derivative of the running coupling constant:
\be 
\beta[g^2(E)] \ =\ \frac {dg^2(E)}{d \ln E}.
\ee
The first terms of the expansion of $\beta(t)$ are
\be
\lb{betaexp}
\beta(t) \ =\ - \frac {b_0 t^2}{8\pi^2} - \frac {b_1 t^3}{128 \pi^4} + \cdots
\ee
with \cite{2 loops} 
$$
b_1 \ =\ \frac {34}3 N_c^2 - \left(   \frac {13}3 N_c - \frac 1{N_c} \right) N_f.
$$

The absence 
of scale invariance at the quantum level means  that the trace
$\hat{\Theta}^\mu_\mu$ of the Heisenberg operator
 describing the
energy-momentum tensor
 is not zero anymore. Accurately  defining what
the operator $\hat{\Theta}^{\mu\nu}$ actually means, one can derive:
 \be
\label{confanom}
\hat{\Theta}^\mu_\mu \ = \ \frac {\beta(g^2)}{2g^4} 
{\rm Tr} \{\hat F_{\mu\nu} \hat F^{\mu\nu} \}.
 \ee
This is the renowned {\it conformal anomaly}.
 It means that the 
classically conserved dilatation
 current (\ref{curdil}) is no longer
 conserved in  the full quantum theory. Conformal symmetry is broken 
explicitly by  quantum effects. Eq.~(\ref{confanom}) is 
an operator equality, i.e.\ all matrix elements of the operators on the left 
and  right hand sides, taken between some physical states,
coincide.

 We will  derive  Eq.~(\ref{confanom}) with path integral methods.
Consider a regularized Euclidean Yang--Mills path integral depending on
an ultraviolet regulator $\Lambda_0$ and a bare coupling constant $g_0^2$. The scale transformation
 affects only the regulator: $\delta \Lambda_{UV} = 
\alpha \Lambda_{UV}$. Due to Eq. (\ref{dimtranexact}), this brings about the same 
change of all physical mass scales $\sim \Lambda_{\rm QCD}$ as the shift of
the bare coupling constant $\delta g_0^2 = -\alpha \beta(g_0^2)$, with 
$\Lambda_{UV}$ kept fixed. The corresponding modification of the
integrand 
$$\exp\{-S^E\} \ = \ \exp\left\{-(1/2g_0^2)
\int {\rm Tr} \{\hat F_{\mu\nu}\hat F_{\mu\nu}\} d^4x\right\} $$
 in the path
integral is then described by the shift
  \be
\label{delSE}
\delta S^E \ =\ 
 \frac{\alpha \beta(g_0^2)
}{g_0^2} S^E . 
\ee
On the other hand, the variation of the Minkowski action is
\be
\lb{sign-theta}
\delta S^M/\delta \alpha \ = \ \int (\pd_\mu J^\mu_D) d^4x = \int \Theta_\mu^\mu d^4x.
\ee
Comparing this with the Minkowski counterpart of Eq. (\ref{delSE}), we 
derive\footnote{A note for pundits:
in supersymmetric
gauge theories, the most natural definition of the Gell-Mann-Low function 
 entering  Eq.~(\ref{intconfan})   is such that it  involves 
only the leading term and the higher-order corrections vanish.
 This makes the expression for the
conformal anomaly
 (\ref{confanom}) 
similar to that of  the {\it chiral} anomaly, Eq. (\ref{chiranom}), which we will  discuss
below. (The coefficient in 
Eq.~(\ref{chiranom}) has a geometric interpretation and does not involve
a series in $g^2$.)
This is not accidental. In supersymmetric
 theories,  the chiral current and the dilatation
 current belong to the same supermultiplet and their anomalies  
are intimately related to each other.}.  
  \be
\label{intconfan}
\int \hat{\Theta}_\mu^\mu d^4x \ =\ \frac{ \beta(g_0^2)}
{2g_0^4} \int {\rm Tr} \{\hat F_{\mu\nu} \hat F^{\mu\nu} \} d^4x  . 
 \ee

Strictly 
speaking, this alone does not guarantee  
 that the  corresponding  {\it integrands}
 also coincide. To show this, one has to find the variation of the path
integral under a {\it local} scale transformation
 (\ref{locCT}). We should
imagine a lattice whose spacing (in physical units) depends on $x$:
$a(x) = a_0[1 + \alpha(x)]$. Somewhat heuristically, 
we might 
say that a theory with  constant
coupling $g_0^2$  defined on the lattice with  $x$-dependent spacing describes   the same physics at   distances that  are much larger
than $a(x)$ as the theory with  $x$-dependent coupling constant $g^2(x) = g^2_0 + \alpha(x) \beta(g^2_0)$ defined on the lattice with constant 
spacing.
Thereby, the variation of the action is
   \be
\label{varYMlocon}
\delta S^E \ =\ \frac{ \beta(g_0^2)
}{2g_0^4} \int \alpha(x)\ {\rm Tr} \{\hat F_{\mu\nu} \hat F_{\mu\nu} \} d^4x .
 \ee
Varying it over $\alpha(x)$, we derive Eq.~(\ref{confanom}).

The anomaly \p{confanom} could also be derived by operator methods, which we will not tackle
here, but illustrate in the following section devoted to:

\section{Anomalous chiral symmetry.}

\setcounter{equation}0
Consider
 Yang--Mills theory with just one massless quark.
The term $i\bar{\psi}/ \!\!\!\!{\cal D} \psi$ in the Lagrangian is invariant
under  global chiral transformations\,\footnote{We will use in the following the spinor representation where 
\be
  \label{convgam}
  \gamma^0 \ =\ \left( \begin{array}{cc} 0 & 1\!\!1 \\ 1\!\!1 & 0 \end{array} \right), \ \ \ \ 
  \gamma^j \ =\ \left( \begin{array}{cc} 0 & - \sigma_j \\ \sigma_j & 0 
\end{array} \right), 
\ee
and
\be
\lb{gamma5}
 \gamma^5 \ \stackrel{\rm def}{=} i\gamma^0\gamma^1\gamma^2\gamma^3\ =\ 
 \ \left( \begin{array}{cc} 1\!\!1 & 0 \\ 0 & -1\!\!1 \end{array} \right). 
\ee
 }
  \be
\delta \psi \ =\ - i\alpha\gamma^5 \psi,\ \ \ \ \ \delta \bar\psi \ = \ -i\alpha
\bar\psi \gamma^5
\label{U1chir}.
 \ee
The corresponding canonical Noether
 current,
 \be
\label{axcur}
 j^{\mu 5} \ =\ \bar\psi \gamma^\mu \gamma^5 \psi,
  \ee
is conserved upon applying the equations of motion.

The Lagrangian is also invariant under the symmetry $\delta \psi = i\beta \psi$,
but it is just a special case of gauge symmetry which, as was mentioned, is
not, in effect, a symmetry and is not a subject of this review.

An important fact is that the symmetry (\ref{U1chir}) exists 
only in the classical case. 
The full quantum path integral is {\it not} invariant under
the transformations (\ref{U1chir}). Like it was also the case with the 
conformal symmetry, this symmetry breaking due to quantum effects can be expressed as an   operator identity involving an anomalous divergence:
  \be
\label{chiranom}
\partial_\mu j^{\mu 5} \ =\ - \frac 1{16\pi^2}
\varepsilon^{\alpha\beta\mu\nu} {\rm Tr} \{\hat F_{\alpha\beta} \hat F_{\mu\nu}\}.
  \ee
There are  many ways to derive and  understand this relation. 
Historically, it was first derived by purely diagrammatic methods. 
 Here we will concentrate on two other ways
which are  more modern and  more general. 
First, we derive (\ref{chiranom}) as an operator 
equality (we skipped an analogous derivation 
when discussing the conformal
anomaly)
 and, second, we will show that the path integral measure is 
actually
not invariant under  chiral transformations, but is modified in such a way
 that the relation (\ref{chiranom}) is satisfied.

Let us start with the operator derivation and, for simplicity, 
consider first 
 the Abelian case. To begin with, we need  to define 
a quantum operator corresponding to the classical axial current
 (\ref{axcur}). The problem is that one cannot harmlessly multiply 
field operators $\psi(x)$ at  coinciding points. Indeed, e.g.\ the vacuum
average $\langle  \psi(x) \bar  \psi(0) \rangle_0 $ behaves as 
$ i/\!\!\!{x}/(2\pi^2 x^4)$  at small $x$, and the
limit $x \to 0$ is singular.
 
\begin{figure}
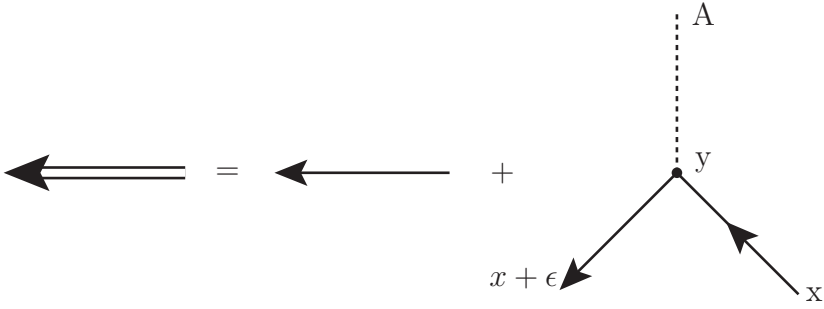

  \begin{axopicture}(120,120)(0,-10)
\SetScale{2}
\Line[arrow,double, arrowpos=1](27,20)(-3,20)
\Text(35,20){=} 
\Line[arrow, arrowpos=1](77,20)(47,20)
\Text(87,20){+}
\Line[arrow, arrowpos=1](120,20)(100,0)
 \Vertex(120,20){1}
\Line[arrow](143,-3)(120,20)
\DashLine(120,20)(120, 50){1}
\Text(125,50){A}
\Text(146,-3){x}
\Text(91,0){$x + \epsilon$}
\Text(125,22){y}
  \end{axopicture}
 \caption{Fermion propagator in an external field.}
  \label{Greenfig}
    \end{figure}

Following Schwinger, we {\it define} the axial current operator of QED
as
 \be
\label{axSchw}
j^{\mu 5} = \lim_{\epsilon \to 0} \bar\psi(x+\epsilon) \gamma^\mu \gamma^5
\exp \left\{ i \int_{x}^{x+\epsilon} A_\nu(y) dy^\nu
 \right\} \psi(x). 
\ee
The factor $\exp \left\{ i \int_{x}^{x+\epsilon} A_\nu(y) 
dy^\nu \right\}$ is very important and makes the expression gauge invariant
in spite of the different arguments of $\bar \psi (x + \epsilon)$ and
 $ \psi (x )$.
 The limit $\epsilon \to 0$ is taken  assuming
averaging over the directions of $\epsilon^\mu$ (otherwise the current 
(\ref{axSchw})  would not
be a Lorentz vector). 
Expanding the exponential up to  terms $O(\epsilon)$, differentiating 
the whole expression (\ref{axSchw}) with respect to $x^\mu$, and making use of the
operator equations of motion $/\!\!\!\!{\cal D} \psi = \gamma^\mu(\partial_\mu
- iA_\mu) \psi = 0$, we obtain
 \be
\label{dmuJmu5}
&&\partial_\mu j^{\mu 5}\ =\ \lim_{\epsilon \to 0} \bar\psi(x+\epsilon) 
\left[ -i\gamma^\mu A_\mu(x + \epsilon)  + i\gamma^\mu A_\mu(x)
 \right. \hspace{2cm}\nonumber \\
  && \left.+ i\epsilon^\nu \gamma^\mu \partial_\mu A_\nu (x) \right] 
  \gamma^5 \psi(x) = 
\lim_{\epsilon \to 0} i \epsilon^\nu F_{\mu\nu}(x) \bar\psi(x+\epsilon) 
 \gamma^\mu  \gamma^5  \psi(x).
 \ee
Superficially, it seems to be zero in the limit $\epsilon \to 0$. This
is not the case, however.
Let us  average Eq.~(\ref{dmuJmu5}) over a state involving a 
classical background 
field $A_\mu(x)$. The fermion Green's function is the free Green's 
function plus the term describing one insertion of the external field plus
 terms with multiple insertions (see  Fig.\  
\ref{Greenfig}). The free Green's function $\sim 
/\!\!\!{\epsilon}/\epsilon^4$ does not
contribute in our case because the corresponding spinor trace ${\rm Tr}
\{\gamma^\alpha \gamma^\mu \gamma^5\}$ is zero in four dimensions. To calculate the graph with one field insertion, it is convenient to choose
the {\it  Fock--Schwinger} \cite{Fock-Schw} (alias, fixed point) gauge
 $(y - x)^\alpha A_\alpha(y) 
= 0$, so that\footnote{The choice \p{fpgauge}  where the potential vanishes at the position $x$ is convenient because it assures  the translational invariance of Green's function. With an arbitrary choice of the fixed point,  it would not be invariant, but the invariance would, of course,  be restored for all  gauge-invariant quantities.}
\be
\label{fpgauge}
A_\alpha(y) = -\frac 12 (y - x )^\beta F_{\alpha \beta}
+ o(y - x )  .
 \ee

An explicit calculation (see  {\bf Problem 1}) gives
 \be
\label{Greenfield}
\!\!\!\!\!\!\!\!\! \langle  \psi(x )  \bar \psi(x + \epsilon) \rangle_A
= - \frac {i /\!\!\! \epsilon}{2\pi^2 \epsilon^4}
-  \frac 1{32\pi^2 \epsilon^2} F_{\alpha\beta} (/\!\!\!\epsilon \gamma^\alpha
\gamma^\beta + \gamma^\alpha\gamma^\beta /\!\!\!\epsilon ) + \ldots \ee
Multiplying this by $-i\epsilon^\nu \gamma^\mu \gamma^5 F_{\mu\nu}$
(the extra minus sign comes from the permutation of anticommuting field
operators), taking the trace,
$${\rm Tr} \{ \gamma^5\gamma^\mu\gamma^\nu\gamma^\alpha \gamma^\beta \} = -4i 
\varepsilon^{\mu\nu\alpha\beta}$$
($\varepsilon^{0123}  = 1$)
and averaging over the directions $\epsilon^\nu\epsilon_\rho/\epsilon^2
 \to \delta^\nu_\rho/4$, we
arrive at the result 
\be
\label{abanom}
\partial_\mu j^{\mu 5} \ =\ - \frac 1{16\pi^2}
\varepsilon^{\alpha\beta\mu\nu} F_{\alpha\beta} F_{\mu\nu} .
  \ee
 Note that the terms describing 
multiple insertions in the Green's function are less singular in $\epsilon$, and their contribution to $\partial_\mu j^{\mu 5}$ vanishes in the limit
$\epsilon \to 0$.

Let us briefly discuss the situation in other even\footnote{If the dimension $D$ is odd,  there is no $\gamma^5$ matrix (the product
$\prod_{\mu = 0}^{D-1} \gamma^\mu$ is proportional to the unit matrix), no axial symmetry and no
anomaly.} dimensions.
In two dimensions, 
 the term with one insertion of the external field 
in the propagator
is $O(\epsilon)$ and does not contribute to the anomaly. But the anomaly
is still there. It appears due to the leading term $ i/\!\!\! 
\epsilon /(2\pi \epsilon^2)$ in the expansion of the fermion Green's function:
in two dimensions, 
\be
\lb{Tr5gg}
{\rm Tr}\{\gamma^\mu\gamma^\nu \gamma^5\} = 2\varepsilon^{\mu\nu} \ \neq 0 
 \ee 
with the convention  
\be
\lb{2Dgamma}
\gamma^0 = \sigma_1 = \left( \begin{array}{cc} 0 & 1 \\ 1 & 0 \end{array} 
\right),\ 
\gamma^1 = i\sigma_2 = \left( \begin{array}{cc} 0 & 1 \\ -1 & 0 \end{array} 
\right),\ 
\gamma^5 = \sigma_3 = \left( \begin{array}{cc} 1 & 0 \\ 0 & -1 \end{array} 
\right),
   \ee
$\varepsilon^{01} = - \varepsilon_{01} =1.$
Eq. \p{dmuJmu5} then  gives:\footnote{The negative sign in the RHS depends on the chosen conventions. With other definitions of $\gamma^5$ and/or of the covariant derivative, the sign may be positive --- see e.g. Eq. (19.18) in Ref. \cite{Pes}.}
 \be
\label{anomd2}
\partial_\mu j^{\mu 5} = - \frac 1{2\pi} \varepsilon^{\mu\nu} F_{\mu\nu}.
  \ee
In six dimensions, both the leading
term and the term linear in  the external field do not contribute because the 
corresponding spinor traces vanish. The anomaly is there due to the
term with two field insertions. For $d = 8$, the term with three insertions
works, etc. In general, one can derive
 \be
\label{anomd}
\partial_\mu j^{\mu 5} =  \pm  \frac 2{n!(4\pi)^n} 
\varepsilon^{\mu_1 \ldots \mu_{2n}} F_{\mu_1\mu_2}
\cdots  F_{\mu_{2n-1}\mu_{2n}},
 \ee
where $d = 2n$. 

\vspace{1mm}

Let us return to  four-dimensional QCD, however. 
We again define  the axial current as in Eq.~(\ref{axSchw}) and the whole 
derivation can be repeated. The only difference is that, along with the terms coming from differentiating the fermion fields in \p{axSchw}, we
also have  to take into account the term $\sim O(\epsilon)$ in the expansion
of the exponential. That will give the contribution
 \be
&&\Delta \left(  \partial_\mu j^{\mu 5} \right)^{\rm non-ab} \ =\ \lim_{\epsilon \to 0}  \bar\psi(x+\epsilon)  \gamma^\mu \gamma^5 \left[\hat A_\mu(x) \int_{x}^{x+\epsilon} \hat A_\nu(y) dy^\nu  \right. \nn
&&\left. - \left( \int_{x}^{x+\epsilon} \hat A_\nu(y) dy^\nu \right) \hat A_\mu(x) \right]\psi(x) 
= \ 
\lim_{\epsilon \to 0} \epsilon^\nu \bar\psi(x+\epsilon) \gamma^\mu \gamma^5
[\hat A_\mu(x), \hat A_\nu(x) ] \psi(x),
\ee
which, being combined with the other terms, just gives the non-Abelian field
strength tensor. Also the relation (\ref{Greenfield}) still holds with the 
full  non-Abelian
$F_{\alpha\beta}$, which is a part  of the magics of the Fock--Schwinger gauge
method. The only
distinction is that the quark Green's function is now a nontrivial matrix not only 
 in spinor, but also in color indices. Thereby, the result 
(\ref{chiranom}) is reproduced.

\vspace{1mm}

Let us now derive the anomaly relation (\ref{chiranom}) with  path
integral methods \cite{Fujikawa}. As we have seen, the anomaly appears due to the necessity
to regularize  theory in the ultraviolet. The  politically most correct
 approach would be to study a path integral regularized
by a lattice. However, an accurate definition of the fermionic action and 
fermionic path integral on the lattice is not so easy. We will 
address this last issue in Sect. 6 and, for the time being, 
use instead  
the more  habitual {\it finite mode}  regularization. 

Consider an Euclidean path integral for the partition function for QCD
with one massless quark flavor. The fermionic part of the integral is
 \be
\lb{fermint-E}
\int \prod_x {d\bar \psi(x) d\psi(x)} \exp\left\{i\int d^4x \bar \psi
 \, /\!\!\!\!{
\cal D}^E \psi  \right\}
\ee
with Hermitian\footnote{
We use the convention where  Euclidean $\gamma$ matrices are  anti-Hermitian,  satisfying 
 \be
\lb{Eucl-com}
 \gamma_\mu^E  \gamma_\nu^E   + \gamma_\nu^E  \gamma_\mu^E  = -2 
\delta_{\mu\nu}.
 \ee
The matrix $\gamma^5$, defined as $\gamma^5 = \gamma_0^E  \gamma_1^E  \gamma_2^E  \gamma_3^E$, is the same as in
the Minkowski space so that 
 \be
\lb{Tr5E}
{\rm Tr}\{\gamma^5 \gamma_\mu^E  \gamma_\nu^E  \gamma_\alpha^E  \gamma_\beta^E\}\  = -4\varepsilon_{\mu\nu\alpha\beta}, \qquad\qquad   \varepsilon_{0123} =1.
 \ee
The Euclidean fermion Lagrangian is ${\cal L}_E  = -i\bar \psi
 \, /\!\!\!\!{
\cal D}^E \psi$.}   ${\cal D}^E $.

An important remark is that the fermion variables $\psi$ and $\bar \psi$ are {\it not} assumed here to be related as $\bar\psi = \psi^\dagger \gamma^0$, as was the case for the Heisenberg field operators in the Minkowski space. Neither they are assumed to be complex conjugate to one another. They represent {\it independent} Grassmann variables in the path integral 
 \p{fermint-E} giving   the determinant of the Dirac operator, 
$\det(-i\,/\!\!\!\!{ \cal D}^E)$.

Let us assume that the theory is somehow regularized in the infrared so 
that the spectrum of the operator $ /\!\!\!\!{ \cal D}^E$ is discrete. 
A proper way to think about the relationship between $\bar \psi(x)$ and $\psi(x)$ is to expand $\psi(x),\ \bar \psi(x)$ over the eigenstates of $ /\!\!\!\!{ \cal D}^E$:
\be
\label{ferEbasis}
\left\{
\begin{array}{c}
 \psi(x) \ =\ \sum_k c_k u_k(x)  \\
\bar  \psi(x) \ =\ \sum_k \bar c_k u_k^\dagger(x)
\end{array} \right. 
\ee
where $\{ u_k(x) \}$ is a complete basis in the corresponding Hilbert space.  It is conveniently chosen as a set of eigenfunctions of the Dirac operator: $ /\!\!\!\!{ \cal D}^E u_k(x) = \lambda_k u_k(x)$; $c_k$ and $\bar c_k$ are independent Grassmann integration variables.
 Then
  \be
\label{dcn}
 \prod_x {d\bar \psi(x) d\psi(x)} \ \stackrel {\rm def}= \ \prod_k d \bar c_k dc_k. 
\ee

Suppose that the field variables are transformed by an infinitesimal
global chiral transformation (\ref{U1chir}).
Now, $\psi' = \psi + \delta \psi$ and  
$\bar \psi' = \bar \psi + \delta \bar \psi$
 can again be expanded in the series (\ref{ferEbasis}). The
new expansion coefficients are related to the old ones:
  \be
\label{cnprim}
c_k' \,=\,c_k  - i\alpha  R_{km} c_m , \qquad \bar c_k' \,=\,\bar c_k - i\alpha  \bar c_m  R_{mk}
 \ee
with 
\be
\lb{Akm}
R_{km} \ =\ \int d^4x \ u_k^\dagger (x) \gamma^5  u_m(x).
 \ee

The point is that the transformation (\ref{cnprim}) has a nonzero Jacobian. 
We have\footnote{The Jacobian appears in the denominator due to the Berezin rule
\be
\lb{Berezin}
\int \prod_k d\bar c_k dc_k \,e^{-\bar c R c} \ =\ \det(R)
\ee
 and the requirement for the integral to stay invariant under a variable change.}
 \be
\label{JacC}
 \prod_k {d\bar c_k}' dc_k' \ =\ J^{-2}  \prod_k d\bar c_k dc_k, 
\ee
where
\be
\label{JviaTr}
J \ =\ {\rm det} (1 - i \alpha R) \ \approx \ 
\exp \left\{ -i \alpha \sum_k \hat R_{kk} \right\}.
 \ee
Or in other words
 \be
\label{lnJ}
\ln J \ =\ - i\alpha \int d^4x \sum_k u^\dagger_k(x) \gamma^5 u_k(x) 
\ + o(\alpha).
 \ee
One's first (wrong!) impression might be that  $ \int d^4x 
\sum_k u^\dagger_k(x) \gamma^5 u_k(x) $ is just zero. Indeed, 
$\{/\!\!\!\!{ \cal D}^E , \gamma^5\} = 0 $ and 
the function $u_k' = \gamma^5 u_k$ is also an eigenfunction of the
Dirac operator  with
  eigenvalue $-\lambda_k$. {\it If} $\lambda_k \neq 0$, $u_k(x)$ and 
$\gamma^5
u_k(x)$ thereby represent  {\it different} eigenfunctions and the integral
$ \int d^4x \ u^\dagger_k(x) \gamma^5 u_k(x) $ vanishes.

A nonzero value of (\ref{lnJ}) is due to the fact that, for intricate
enough {\it topologically nontrivial} gauge fields, the spectrum of the 
Dirac operator involves some number 
of exact zero modes
 for which $\gamma^5 
u_0(x) = \pm u_0(x)$ (depending on whether the modes are right-handed
or left-handed) and their contribution in the sum (\ref{lnJ}) is 
responsible for the whole effect. 
A famous theorem of Atiyah and Singer \cite{AS}
 dictates:
  \be
\label{AtSing}
 \int d^4x \sum_k u^\dagger_k(x) \gamma^5 u_k(x)  = n_R^{(0)} -
n_L^{(0)} \ =\ q ,
 \ee
where $n_{R,L}^{(0)}$ is the number of the right-handed (left-handed)
zero modes
 and $q$ is the {\it Pontryagin index} of the gauge
field configuration:\footnote{The most familiar configurations with nonzero Pontryagin index $q = 1$ are BPST instantons \cite{BPST}:
 \be
    \label{Ainst}
   A^a_\mu = \frac{2\eta_{\mu\nu}^a x_\nu}{x^2 + \rho^2},
\ee
where $\eta_{\mu\nu}^a$ are 
the so-called 't Hooft symbols:
 \be
 \label{eta}
 \eta_{44}^a = 0\ ,\ \ \eta_{jk}^a = \epsilon_{ajk}\ ,\ \ 
 \eta_{4j}^a = -\delta_{aj}\ ,\ \  \eta_{j4}^a = \delta_{aj}.
 \ee
The field strength correponding to the gauge potential \p{Ainst} is
\be
   \label{Finst}
  F_{\mu\nu}^a =  - \frac{4\rho^2 \eta_{\mu\nu}^a}{(x^2 + \rho^2)^2}.
 \ee}
 \be
    \label{k4F}
     q \ =\ \frac 1{16\pi^2}  \int
   d^4x \ {\rm Tr} \{\hat F_{\mu\nu} \hat{\tilde F}_{\mu\nu}\},
 \ee 
   where $ \hat{\tilde F}_{\mu\nu} = \frac 12
   \varepsilon_{\mu\nu\alpha\beta} \hat F_{\alpha\beta}$.

\vspace{1mm}

How to prove this theorem? The mathematical proof due to Atiyah and Singer is complicated and is not familiar to your author.  We present here a proof, which is adapted to the problem in interest and relies solely on the standard apparatus of quantum mechnics. Consider first a little  more simple case where the gauge field is Abelian.

 As all nonzero modes  contribute  zero to the 
sum, we are allowed to consider instead of (\ref{AtSing}) the sum
 \be
\label{sumM}
 \int d^4x \sum_k u^\dagger_k(x) \gamma^5 u_k(x) e^{-\lambda_k^2/M^2}.
 \ee
The finite parameter $M$ brings about the finite mode regularization
 mentioned above:
the contribution of  modes with large $\lambda_k^2$ is suppressed.
To calculate the sum, consider a quantum mechanical problem with
the matrix Hamiltonian\footnote{We follow the tradition and use the notation $\hat X$  for two rather different purposes. It may mark the color matrix nature of $\hat X = X^a t^a$ as for $\hat A_\mu$ or its operator nature as for $\hat H$ and $\hat p_\mu$ in Eq. \p{oper H}. Hopefully the reader will not be confused by that.} 
\be 
\lb{oper H}
\hat H = (/\!\!\!\!{\cal D})^2 = - [\gamma_\mu^E(\hat p_\mu - A_\mu)]^2, 
\ \ \ \ \hat{p}_\mu  = -i\partial_\mu . 
\ee
 The sum (\ref{sumM}) is nothing but 
$\int d^4x \, {\rm Tr} \{\gamma^5 {\cal K}(x,x;1/M^2)\}$, where ${\cal K}$ is
the  evolution operator
 of our quantum mechanical system with 
imaginary time $\beta = 1/M^2$. It is also a matrix.
[Note that the phase space of our system is
 8-dimensional, involving the
coordinates $x_\mu, p_\mu$, and the evolution occurs in some unphysical
``fifth'' time.] The trace is done over the spinor and color indices.

The Euclidean evolution operator
 can be expressed as a path integral, 
\be
{\cal K}(x,x; \beta) \ =\ \int \exp\left\{i \int_0^\beta \left[p_\mu \frac {dx_\mu}{d\tau}  + i  H(p,x) \right] d\tau \right\}
\prod_\tau \frac {dp_\mu(\tau) dx_\mu(\tau) }{2\pi}
 \ee
with the classical (but still spinor matrix) $H(p,x)$.
It runs over the trajectories $x_\mu(\tau)$ and $p_\mu(\tau)$ satisfying the periodic boundary conditions
\be
\lb{bc}
 x_\mu(\beta) = x_\mu(0) = x, \qquad p_\mu(\beta) = p_\mu(0).
 \ee
 Assume now that 
$M$ is very large, so that $\beta$ is very small. Bearing in mind \p{bc}, $x_\mu(\tau)$ and $p_\mu(\tau)$ stay practically constant, the term  $\propto  {dx_\mu}/{d\tau}$ in the exponent can be neglected and the functional integral reduces to an ordinary 
finite-dimensional integral over momenta:
\be
{\cal K}(x,x; \beta) \ \stackrel{{\rm small}\,  \beta}{\approx} \  \int \frac {d^4p}{(2\pi)^4} e^{-\beta  H(p,x)} .
\ee 
 We derive:
  \be
\label{intCZ}
\int d^4x \, {\rm Tr} \{\gamma^5 {\cal K}(x,x;1/M^2)\} \ \approx \ 
\int \frac {d^4x d^4p}{(2\pi)^4}
 {\rm Tr} \left\{ \gamma^5 e^{-(/\!\!\!\!{\cal D})^2/M^2} \right\} . 
 \ee
Using the convention \p{Eucl-com}, giving $$(/\!\!\!\!{\cal D})^2 \ =\  - {\cal D}^2 -  (i/2) 
\gamma^E_\mu \gamma^E_\nu F_{\mu\nu}  = (p_\mu - A_\mu)^2 -  (i/2) 
\gamma^E_\mu \gamma^E_\nu F_{\mu\nu},  $$ 
evaluating the trace using  \p{Tr5E}
and shifting the variable of the integration $p_\mu - A_\mu  \to  p_\mu$, the integral can be easily calculated
to  leading order in $1/M^2$. The result is the Abelian version of $q$:
 \be
q_{Ab} \ =\  \frac 1 {16\pi^2}\int
   d^4x  \,  F_{\mu\nu} {\tilde F}_{\mu\nu}.
 \ee
The same reasoning works also in the non-Abelian case with the only  complication 
 that $(p_\mu - A_\mu)^2$ is now replaced by a color matrix $(p_\mu - A^a_\mu t^a)^2$, and one cannot just shift  the momentum variable in the integral.  

But actually one can. To see that, consider the integral $\int_{-\infty}^\infty  \exp\{-(p - \hat A)^2 \}$, where $\hat A$ is some $p$-independent matrix, and expand it in $\hat A$. We obtain
 $$ \int_{-\infty}^\infty dp\, \exp\{-(p - \hat A)^2 \} \ =\ \int_{-\infty}^\infty dp\,  e^{-p^2} - \hat A  \int_{-\infty}^\infty dp \left[\frac \pd{\pd p}  e^{-p^2} \right] + \frac 12 \hat A^2 \int_{-\infty}^\infty dp \left[\frac {\pd^2}{\pd^2 p}  e^{-p^2} \right] +\cdots
$$ 
All the terms besides the first one vanish, and the integral does not depend on $\hat A$.
Thus, the index theorem (\ref{AtSing}) 
is proven.\footnote{By the way,  the Atiyah--Singer
 index for
the Dirac operator is identical  to the  {\it Witten index}
 of a certain  supersymmetric
 quantum-mechanical system 
(with an opposite sign for the manifolds of dimension $4k+2$ ), but we will not delve into further details here addressing an interested reader to the original papers \cite{SUSY-AS} and to  Chapter 15 of the book \cite{glasses}.) 
The  method for deriving Eq.~(\ref{AtSing}) presented above coincides in fact 
with the known derivation of an integral representation 
for  the Witten index due to Cecotti and Girardello. \cite{Cecotti}}

\vspace{1mm}

Substituting \p{JviaTr} into \p{JacC}, with bearing in mind \p{AtSing} and \p{k4F}, we now see that the change 
 of the measure under a chiral transformation
can be represented as a shift of the effective Euclidean action
  \be
\label{althet}
\delta S^E \  = \ -\frac {i\alpha}{16\pi^2} \epsilon_{\alpha\beta\mu\nu} 
\int d^4x {\rm Tr} \{ \hat F_{\alpha\beta} \hat F_{\mu\nu}
 \}. \ee
The shift of the  Minkowski action is given by the same expression 
without the prefactor  $i$.
In other words, a global chiral transfomation 
is equivalent to leaving the fermionic fields intact, but  shifting instead
the parameter $\theta$ in the original theory \p{LQCD}:
$\theta \to \theta - 2\alpha$. 

Just like in the case of the conformal anomaly,
 we have studied up to now only
the change of the measure under  global symmetry transformations. In the
chiral symmetry case, it is not too difficult to find out how the
measure  changes under  local transformations with  $x$-dependent
parameters $\alpha(x)$. We have
   \be
\label{lnJloc}
\ln J[\alpha(x)] \ =\ -i \int d^4x \ \alpha(x)\ \sum_k u^\dagger_k(x) \gamma^5 
u_k(x) \ + o(\alpha). 
\ee  
Repeating all the steps of the  above derivation, 
we find\footnote{Note that this is now a {\it local} quantity not directly
 related to
the global properties of a gauge field configuration like a nonzero net 
topological charge $q$ and
the presence of fermion zero
 modes.}
$$
\lim_{M \to \infty}  \sum_k u^\dagger_k(x) \gamma^5 u_k(x) 
e^{-\lambda_k^2/M^2} \ =\ 
\frac 1{32\pi^2}
\epsilon_{\alpha\beta\mu\nu} {\rm Tr} \{ \hat F_{\alpha\beta} \hat F_{\mu\nu} \}(x)
,
$$
which gives
    \be
\label{allocthet}
\delta S^E \ = \ -\frac {i}{16\pi^2} \epsilon_{\alpha\beta\mu\nu} 
\int d^4x \ \alpha(x) {\rm Tr} \{ {\hat F}_{\alpha\beta} \hat F_{\mu\nu} \} (x)
 . \ee
Varying the corresponding shift of the Minkowski action with respect to $\alpha(x)$, we obtain the anomalous
 divergence of the 
axial current in accordance with (\ref{chiranom}).

The last  comment is that, in  real QCD with several light quarks, each
flavor contributes on  equal footing to the anomaly of the singlet
axial current $j^{\mu 5 ({\rm singl})} = \sum_f \bar\psi_f \gamma^\mu \gamma^5
\psi_f $, and the  result
(\ref{chiranom}) is just multiplied by $N_f$.

\vspace{.4cm}

{\bf Problem 1}. Using the Fock--Schwinger
 gauge (\ref{fpgauge}), derive the
expression (\ref{Greenfield}) for the fermion Green's function.

\vspace{.1cm}

{\bf Solution}. 
With the gauge choice (\ref{fpgauge}), the Green's function depends only on the
difference  of coordinates $x - (x+\epsilon) = -\epsilon$. We can set $x = 0$ 
and write
 \be
\label{GAx}
\langle  \psi(0 )  \bar \psi( \epsilon) \rangle_A = 
G_0(-\epsilon) + \int d^4y \, G_0(y - \epsilon) i \gamma^\alpha A_\alpha (y ) 
G_0(-y ),
 \ee
where $G_0(u)$ is the free fermion Green's function:
$$ G_0(x) \ = \ \langle \psi(x) \bar \psi(0 \rangle_0 \ = \int \frac {d^4p }{(2\pi)^4} e^{-ipx} G(p). $$ 
Substituting here $A_\alpha(y) = - \frac 12 
y^\beta F_{\alpha\beta}$ and going over into  momentum space, we obtain
 \be
\label{GAp}
G_A(p) = \ \frac {i /\!\!\! p}{p^2} - 
\frac {F_{\alpha\beta}}2   \frac {/\!\!\! p}{p^2 }  \gamma^\alpha  \left(  \frac   {\partial}
{\partial p_\beta}   \frac {/\!\!\! p}{p^2 } \right)
 \ = 
\frac {i /\!\!\! p}{p^2} - \frac {F_{\alpha\beta}}{4 p^4} (/\!\!\! p 
\gamma^\alpha \gamma^\beta + \gamma^\alpha \gamma^\beta /\!\!\! p) .
\nonumber\\
 \ee
Its Fourier image gives Eq.~(\ref{Greenfield}). 

\vspace{.4cm}

{\bf Problem 2}.  For the instanton field configuration (\ref{Ainst}), 
show that the function
  \be
 \label{inst0mode}
u_0(x) \ =\ \frac{\rho}{\pi(x^2 + \rho^2)^{3/2}}
\left( \begin{array}{c}  \epsilon_{i\alpha}  \\ 0 \end{array} \right)
 \ee
($i = 1,2$ is the color and $\alpha = 1,2$ is the spinor index)
represents
an exact normalized right-handed zero mode
 of the Dirac equation  \cite{Hooft}).

\vspace{.1cm}

{\bf Solution}. It is convenient to express the Euclidean $\gamma$ matrices
 as
\be
 \label{gamEsig}
\gamma_\mu^E \ =\ \left( \begin{array}{cc}
  0 & - \sigma_\mu^\dagger \\
\sigma_\mu & 0 \end{array} \right)
 \ee
with $\sigma_\mu = (i, \vecg{ \sigma} )$. It is easy to check that such matrices satisfy the properties \p{Eucl-com} and \p{Tr5E}.

 The matrices $\sigma_\mu$ satisfy the
relations
  \be
 \label{sigmurel}
\sigma_\mu^\dagger \sigma_\nu + \sigma_\nu^\dagger \sigma_\mu &=&
\sigma_\mu \sigma_\nu^\dagger + \sigma_\nu \sigma_\mu^\dagger \ =\ 
2\delta_{\mu\nu} \ ,\nonumber \\ 
\sigma_\mu^\dagger \sigma_\nu - \sigma_\nu^\dagger \sigma_\mu &=&
2i \eta^a_{\mu\nu} \sigma^a \ , \nonumber \\
  \sigma_\mu \sigma_\nu^\dagger - \sigma_\nu \sigma_\mu^\dagger &=&
2i \bar \eta^a_{\mu\nu} \sigma^a.
  \ee
Note also the corollaries $$\sigma^\dagger_\mu \sigma_\nu \sigma^\dagger_\mu
= -2  \sigma_\nu^\dagger\ ,\ \ \ \sigma_\mu \sigma_\nu^\dagger \sigma_\mu
= -2  \sigma_\nu \ .$$
Let us first  convince ourselves  that  only the right-handed solutions [satisfying $\gamma^5 u = u$ with $\gamma^5$ =   diag $(\mathbb{1}, -\mathbb{1})$,  as defined in Eq. \p{gamma5}]  to 
the 
equation $/\!\!\!\!{\cal D}^E u = 0$ are admissible. Indeed, a left-handed
solution should satisfy 
$$   /\!\!\!\!{\cal D}^E u_L \ =\ \gamma^E_\mu {\cal D}_\mu \left(\begin{array}{c} 0 \\ u_L \end{array}  \right) \ =\ 0  
$$
giving 
$\sigma_\mu^\dagger {\cal D}_\mu u_L = 0$.  Act upon 
this with the operator  $\sigma_\nu {\cal D}_\nu$. We obtain
  \be 
\label{actL}
 \sigma_\nu {\cal D}_\nu \sigma_\mu^\dagger {\cal D}_\mu \ &=&\  
( {\cal D}_\mu)^2 + \frac 14 [
\sigma_\nu \sigma_\mu^\dagger - \sigma_\mu \sigma_\nu^\dagger  ]
[{\cal D}_\nu , {\cal D}_\mu ]\hspace{2.2cm} 
\nonumber \\
\!\!\!\!\!\!\! &=& ( {\cal D}_\mu)^2 -
 ( \bar \eta^a_{\nu\mu} \sigma^a)  ( \eta^b_{\nu\mu} \sigma^b) 
\frac {\rho^2}{(x^2 + \rho^2)^2} \ =\ ( {\cal D}_\mu)^2, 
\ee
  where we have substituted the instanton field strength
 $[{\cal D}_\nu , {\cal D}_\mu ] = -i F_{\nu\mu}$ from Eq.~(\ref{Finst})  and used the
property $\bar \eta^a_{\nu\mu} \eta^b_{\nu\mu} = 0$. 
But the operator $-( {\cal D}_\mu)^2\ = \ ( i {\cal D}_\mu)^2 $  is a
 sum of 
squares of Hermitian operators and is thus positive definite. The 
equation ${\cal D}^2 u_L = 0$ and hence the equation 
 $\sigma_\mu^\dagger {\cal D}_\mu u_L = 0$ have no solutions.

For the right-handed spinors, the zero mode equation has the form
 \be
 \label{zeroR}
(\sigma_\mu)_{\alpha\beta} \left[ \partial_\mu  - \ 
\frac {i \eta^a_{\mu\nu} x_\nu \sigma^a }{x^2 + \rho^2}
\right]_{ij} u_{j\beta}^R (x) \ =\ 0
 . \ee
Using the relations (\ref{sigmurel}) and the property $\sigma_\mu^T
= - \sigma_2 \sigma_\mu^\dagger \sigma_2 $, we can rewrite this equation as
 \be
\label{zeroR1}
\left[ \partial_\mu  - \ 
\frac { x_\nu} {2(x^2 + \rho^2)}( \sigma_\mu^\dagger \sigma_\nu - 
\sigma_\nu^\dagger \sigma_\mu ) \right] u \sigma_2 \sigma_\mu^\dagger
 \ =\ 0
  . \ee
The natural ansatz $u_{i\alpha} = (\sigma_2)_{i\alpha} G(x^2)
= -i \epsilon_{i\alpha} G(x^2)$ satisfies the equation provided
$$ 2G'(x^2) + \frac {3 G(x ^2)}{x^2 + \rho^2}  \ =\ 0
\ \ {\rm and\ hence} \ \ \ G(x^2) \ =\ \frac C{(x^2 + \rho^2)^{3/2}}\ .$$
The particular value of $C$ in (\ref{inst0mode}) normalizes the solution
to one: 
 $$\int d^4x \ u_{i\alpha}^\dagger  u_{i\alpha} \ =\ 1\ .$$ 
Due to the index theorem
 (\ref{AtSing}), and since left-handed solutions are
absent, the instanton field does not admit other 
zero mode
 solutions.

\section{Nonsinglet chiral symmetry and its spontaneous  breaking.}

\setcounter{equation}0

A theory with several flavors of massless quarks  enjoys, besides the 
singlet axial
symmetry $\delta \psi_f = - i \alpha \gamma^5 \psi_f$ (which, as we have seen,
it {\it does} not actually enjoy), a set of flavor-nonsinglet symmetries:
  \be 
\label{symvect}
\delta \psi_f =  -i \alpha_a  [t^a \psi]_f
  \ee
and
  \be 
\label{symax}
\delta \psi_f = -i \beta_a \gamma^5 [t^a \psi]_f
  , \ee
where $t^a$ are the generators of the flavor $SU(N_f)$ group normalized in a standard way,
$$ {\rm Tr} \{t^a t^b\} \ = \ \frac 12\delta^{ab}. $$
 The symmetry (\ref{symvect}) is the
 ordinary isotopic symmetry.
It is still present  even if the quarks are endowed with 
a mass (of the same magnitude for all
flavors). The symmetry (\ref{symax}) holds only in  massless theory. The 
corresponding Noether currents
 are
   \be 
(j^\mu)^a \ = \ \bar \psi t^a \gamma^\mu \psi, \ \ \ \ 
(j^{\mu 5})^a
 \ = \ \bar \psi t^a \gamma^\mu \gamma^5 \psi
 \label{curvecax}
  . \ee
They are not anomalous, i.e.\ they are duly conserved not only at
 the classical level, 
but also in  full quantum theory. To describe a finite element of the 
symmetry group, it is convenient to define left-handed and right-handed
quark fields:
\be
  \label{psiLR}
\psi_{L,R}  = \ \frac 12 (1 \mp \gamma^5) \psi\ , \ \ \ \ 
\bar\psi_{L,R}  = \ \frac 12 \bar\psi (1 \pm \gamma^5).
 \ee
One can easily see that the Lagrangian of massless QCD is invariant under
 the transformations
  \be
\label{chirtrans}
\psi_{L} \ \to \ V_L \psi_{L}, \ \ \ \ 
\psi_{R} \ \to \ V_R \psi_{R}, 
  \ee
where $V_L$ and $V_R$ are two different $U(N_f)$ matrices. The singlet
axial transformations with $V_L = V_R^* = e^{i\phi}$ are anomalous
 by the
same token as in the theory with only one quark flavor. Therefore, the true
fermionic symmetry group of  massless QCD is
  \be
\label{SULSUR}
{\cal G} \ = \ SU_L(N_f) \times SU_R(N_f) \times U_V(1)
 . \ee
A fundamental {\it experimental fact} is that the symmetry (\ref{SULSUR})
is actually {\it spontaneously broken}, which means that the vacuum state
 is not invariant under the action of the group ${\cal G}$. The symmetry
${\cal G}$ is, however, not broken completely. The vacuum is still invariant under the transformations  with $V_L = V_R$, 
generated by the vector isotopic 
currents.\footnote{An exact {\it Vafa-Witten theorem} \cite{Vafa} that the vector 
symmetry cannot be
 broken spontaneously in QCD will be proven in Sect. 7.} 
Thereby, the pattern of breaking is
  \be
\label{patbreak}
 SU_L(N_f) \times SU_R(N_f) \ \to  SU_V(N_f)
  . \ee
  Spontaneous breaking of the axial symmetry shows up in the appearance of 
nonzero vacuum expectation values,
 \be
\label{condfg}
\Sigma^{fg}\ = \ \langle  \psi_{L}^f  \bar \psi_{R}^g \rangle_0\ 
 \ee
(the {\it quark condensate}
 matrix).
Nonbreaking of the vector symmetry implies that the matrix order parameter
(\ref{condfg}) can be brought into the form
   \be
\label{conddel}
\Sigma^{fg}\ = \  \frac 12 \Sigma \,\delta^{fg}
 \ee
by the group transformations (\ref{chirtrans}).
This means  that a generic condensate matrix $\Sigma^{fg}$ is a 
unitary $SU(N_f)$ matrix multiplied by $\Sigma$. 

In general, $\Sigma$ could be any complex number. It can be
made real by a global $U_A(1)$ rotation which, according to Eq.~(\ref{althet})
(with the factor $N_f$),
amounts to a shift of the vacuum angle $\theta$. In other words, in the theory
with quarks the
physics does not depend on the parameter $\theta$ and the phase of $\Sigma$
separately, but only on their combination $\theta - N_f {\rm arg} (\Sigma)$.
The earlier mentioned  fact that the experimental value of $\theta$ is very 
small actually refers  to {\it this} particular combination.
 It is convenient then
to choose $\Sigma$ real and positive and $\theta = 0$. 
From  experiment\footnote{From the Gell-Mann--Oaks--Renner relation \p{GMOR}, from QCD sum rules \cite{sumrul}  and from lattice numerical simulations \cite{condlat}. }  we know that  $\Sigma
 \approx (250 \ {\rm MeV})^3$ with about 30\% uncertainty.

By  {\it Nambu--Goldstone theorem} \cite{Nambu}, 
if a global continuous symmetry is broken spontaneously, 
purely massless particles called {\it Goldstone bosons}
 appear in the 
spectrum. The simplest example for this  is the theory of a complex
scalar field $\phi(x)$ with a ``Mexican hat'' potential $\lambda (\bar \phi
\phi - \mu^2)^2$. The Lagrangian is invariant with respect to global phase 
rotations $\phi(x) \to e^{i\alpha} \phi(x) $, but the vacuum state (say, the state with $\langle \phi \rangle_{\rm vac} = \ \mu$) is characterized by a particular
value of the phase playing the role of an order parameter. 
The point is that a field configuration involving 
small fluctuations of the phase $\phi(x)
= \mu e^{i\alpha(x)},\ \alpha(x) \ll 1$, still has  zero potential energy
while the kinetic energy is proportional to 
$\int d^4x (\partial_\mu \alpha)^2$. Quantizing 
this effective Lagrangian gives massless particles. 

In general, the number of Goldstone particles
 is equal to the dimension of the
vector space formed by the generators of the gauge group which are ``broken'',
 i.e.\ act nontrivially  on the vacuum state. By construction, for any 
such generator $J$, a Goldstone branch in the spectrum 
$|\alpha_J \rangle_p$ exists, so that the
matrix elements $\langle 0 | J | \alpha_J \rangle_p$ are not zero. In our case,
the breaking (\ref{patbreak}) is associated with the
appearance of $N_f^2 - 1$ Goldstone particles [$N_f^2 - 1$ is
the dimension of the original group ${\cal G}$ minus the dimension of the
residual group $U_V(N_f)$]. As it is the axial symmetry
which is broken, the Goldstone particles are pseudoscalars. 

An analogy with an ordinary  iron bar (with one of its magnetic domains) is very instructive
 here. The Hamiltonian
of a ferromagnet is rotationally invariant. Spontaneous 
magnetization signals
the spontaneous breaking of the rotational invariance $SO(3)$ down to $SO(2)$.
The direction ${\vec{n}}$ of the magnetization $\langle  \vec{M}  \rangle = M_0
\vec{n}$, with $\vec{n}^2 = 1$, is arbitrary. Rotating the reference
frame, we can choose, say, $\vec{n} = (0,0,1)$ [cf.\ Eq.~(\ref{conddel})].
Fluctuations of the vector $\vec{n}$ in space and time are described by $3-1
= 2$ parameters and correspond to  magnon massless excitations.

\section{Effective chiral lagrangian.}
\setcounter{equation}0

It is a fundamental and important fact  that
spontaneous breaking of a continuous symmetry not only leads to
massless Goldstone particles,
 but also fixes the 
{\it interactions} of the latter at low energies.
To see this, let us first recall  that the Goldstone field describes 
fluctuations of the order parameter: $\Sigma^{fg} \to \Sigma^{fg}(x) =
\frac 12 \Sigma U(x) $ with $U(x) \in SU(N_f)$. It is convenient to express 
$U(x)$ as an exponential
  \be
 \label{Uchirexp}
U(x) \ =\ \exp \left\{  \frac {2i \phi^a(x) t^a}{F_\pi} \right \},
 \ee
where $\phi^a(x)$ are the physical meson fields [so that $\phi^a(x) = 0$ 
corresponds
to the vacuum (\ref{conddel})] and $F_\pi$ is a constant carrying dimension of mass.

In massless QCD, Goldstone particles
 are massless whereas all other states in the physical
spectrum have  nonzero mass. Therefore, we are  in the 
{\it Born--Oppenheimer} situation:
 there are two distinct energy scales and one can  write down an effective Lagrangian depending only on {\it slow} Goldstone fields, 
with the {\it fast} degrees of freedom corresponding to all other particles 
being integrated out.\footnote{In the pioneer paper \cite{BO}, Born and Oppenheimer found the spectrum of the hydrogen molecule  by solving first the Schr\"odinger equation involving fast electron degrees of freedom and analysing then the effective Hamiltonian that depended on the positions and momenta of heavy protons.}

To fix the exact form of this Lagrangian, note that the transformations
(\ref{chirtrans}) are realized at the level of the effective Lagrangian as $U \to V_L U V_R^\dagger$. 
Any scalar function depending on $U$ and invariant
under this symmetry is also a function of $U^\dagger U = 1$, i.e.\ it is just
a constant. There is only one invariant structure involving two derivatives:
 \be
\label{Leffchir2}
{\cal L}_{\rm eff}^{(2)} \ =\ \frac {F_\pi^2}4 {\rm Tr} \left\{ \partial_\mu U
 \partial^\mu U^\dagger \right \}
 . \ee
Take $N_f = 2$. The  perturbative expansion of Eq.~(\ref{Leffchir2}) 
in powers of $\phi$ reads
   \be
\label{Lexpphi}
{\cal L}_{\rm eff}^{(2)} \ =\ \frac 12 ( \partial_\mu \phi^a )^2 +
\frac 1{6 F_\pi^2} \left[ (\phi^a  \partial_\mu \phi^a )^2 - (\phi^a \phi^a)
(\partial_\mu \phi^b)^2 \right] + \cdots
 \ee
Also  for $N_f \geq 3$, we have, on top of the standard kinetic term, a quartic
term involving two derivatives, with somewhat   more 
complicated group structure.
We see that the  symmetry dictates  rather specific interactions between
the Goldstone bosons.
 They do not interact at the $S$ wave level, 
which means that the amplitude of their scattering vanishes at zero momenta, 
but the strength of
interaction grows rapidly with energy. 

Equation (\ref{Leffchir2}) describes the 
effective chiral Lagrangian to leading order. The first corrections 
involve 4 derivatives and there are 3 different invariant functions
   \be
\label{Leffchir4}
{\cal L}_{\rm eff}^{(4)} \ =\ L_1 
 {\rm Tr} \left\{ \partial_\mu U \partial^\mu U^\dagger \right \}^2 
&+& L_2 {\rm Tr} \left\{ \partial_\mu U \partial_\nu U^\dagger \right \}
{\rm Tr} \left\{ \partial^\mu U \partial^\nu U^\dagger \right \} 
\nonumber \\
 &+& \ L_3  {\rm Tr} \left\{ \partial_\mu U \partial^\mu U^\dagger 
 \partial_\nu U \partial^\nu U^\dagger \right\} 
 \ee
(only 2 linearly independent structures are left for $N_f = 2$). 

The relevant Born--Oppenheimer expansion parameter is $$\kappa_{\rm chir}
\sim p^{\rm char}/F_\pi\ .$$ When  $\kappa_{\rm chir} \sim 1$ (in practice, one
should rather take  $\kappa_{\rm chir} \sim 2\pi$), 
the  Born--Oppenheimer approximation,
 as well as  
the whole effective Lagrangian approach,
 breaks down, and
non-Goldstone degrees of freedom become important. The physical
meaning of $F_\pi$ is thus clarified. It characterizes the gap in the
spectrum and sets a scale below which massive degrees of freedom can be 
disregarded.

Let us discuss the real world now. The  Lagrangian of real QCD 
(\ref{LQCD}) is 
not invariant under the axial symmetry transformations just because quarks
have nonzero masses. The symmetry (\ref{SULSUR}) is still very much relevant 
for QCD because {\it some} of the quarks happen to be very light. This is
especially so for $u$ and $d$ quarks whose masses --- $m_u \approx 3$ MeV and  $m_d \approx 5$ MeV ---
 are much smaller than the
characteristic hadron scale $\mu_{\rm hadr} \approx .5 \ {\rm GeV}$:
the symmetry (\ref{SULSUR}) is almost there!

Spontaneous breaking of an exact $SU_L(2) \times SU_R(2)$ symmetry would lead
to the existence of three strictly massless Goldstone bosons. As the symmetry is 
not quite exact, the Goldstone particles
 have a small mass. However, 
their mass $M$ goes
to zero in the chiral limit $m_{u,d} \to 0$.
Indeed, trading the mass term
 \be
- m_u \bar u u - m_d \bar d d = m_u (u_L \bar u_R +  u_R \bar u_L)
+ m_d (d_L \bar d_R +  d_R \bar d_L)
 \ee
in the QCD Lagrangian for the contribution\footnote {Equation (\ref{Lchirmass}) is the leading chiral-noninvariant 
contribution
in ${\cal L}_{\rm eff}$. Also  terms of higher order in ${\cal M}$, as well
as  terms of  first order in ${\cal M}$ but involving derivatives of $U$,
are allowed.} 
 \be
 \label{Lchirmass}
{\cal L}_{\rm eff}^{(m)} \ =\ 
\Sigma \ {\rm Re} \left[ {\rm Tr} \{ {\cal M} U \} \right] 
  \ee
[${\cal M}$  is the quark mass matrix
which is chosen here in the form  ${\cal M} = {\rm diag} (m_u, m_d)$ with real $m_u, m_d$; do not confond ${\cal M}$ with $M$ !] in the effective chiral Lagrangian  and expanding Eq.~(\ref{Lchirmass})
in $\phi^a$, we obtain the 
{\it Gell-Mann--Oakes--Renner relation} \cite{GMOR},
   \be
 \label{GMOR}
F_\pi^2 M^2  =  (m_u + m_d)\Sigma + O(m_q^2)
. \ee
These light pseudo-Goldstone particles
 are well know to experimentalists. They
are nothing but the pions. Experimentally, $F_\pi \approx 93 $ MeV. 
The  constant $F_\pi$ appears also in  the matrix element 
$\langle {\rm vac} | A_\mu^+ |\pi \rangle = i \sqrt{2} F_\pi 
p_\mu^\pi$ of the
axial current $A_\mu^+ = \bar d \gamma_\mu \gamma^5 u$   and determines the 
charged pion decay rate. 
The gap between the pseudo-Goldstone sector and the massive sector in QCD is of order of  the mass of $\rho$ meson,
 $M_\rho \approx 8F_\pi$.

In the real world, there is also a third relatively light quark --- the strange
quark. Its mass, $m_s \approx 100 \ {\rm MeV}$, is still small enough for the 
symmetry (\ref{SULSUR}) to make  sense. Thus, QCD enjoys an approximate
 $SU_L(3) \times SU_R(3)$ symmetry which is broken spontaneously with the 
appearance of the quark condensate
 (\ref{conddel}) and also explicitly
due to nonzero quark masses. As the latter are relatively small, one can
build up a Born--Oppenheimer expansion
 (alias, {\it chiral perturbation theory} \cite{CPT})
over the small parameters $p^{\rm char}/\mu_{\rm hadr}$ and  
$m_q/\mu_{\rm hadr}$. The spectrum of QCD includes
8 light pseudo-Goldstone  mesons,
 the well-known pseudoscalar octet 
$(\pi, K, \eta)$. 

Exploiting the symmetry (\ref{SULSUR}) allows one to obtain many nontrivial
predictions for the properties of these mesons, but this, as well as
many other
 wonderful achievements  in describing the physics of hadrons through QCD,
 is  beyond the scope of our review.

\section{QCD on Euclidean lattice}
\setcounter{equation}0

To attribute meaning to the path integral symbol, one should regularize it --- replace an infinite-dimensional integral by a finite-dimensional one. The most natural way to do so is making space-time discrete and replace a continuum of points on $\mathbb{R}^4$ by a set of nodes of hypercubic lattice. A finite distance $a$ between adjacent nodes provides ultraviolet regularization. A finite size $L$  of this lattice provides infrared regularization. The properties of physically interesting continuum field theory can be inferred  by studying the limit $a \to 0$, $L \to \infty$.  It is all possible to do in Euclidean space where the measure in the path integral $\sim \exp\{-S^E\}$ is positive definite and the lattice integral is well defined. In Minkowski space with the measure
  $\sim \exp\{iS^M\}$, a rigorous definition of path integral that would satisfy mathematicians is not available though some practical numerical calculations are still possible \cite{GE-Blin}. In our review, we will stay in Euclidean space, however.

We describe first how to define path integral on Euclidean lattice for pure Yang--Mills theory while keeping gauge invariance. Then we include fermions and tackle the problem of chiral symmetry in the lattice version of QCD. Our report will necessarily be rather succinct. A reader who wants to learn it in more details is invited to consult an excellent book \cite{Gatt}. 

\subsection{Pure gauge theory}

\begin{figure} [ht!]
      \bc
    \includegraphics[width=.5\textwidth]{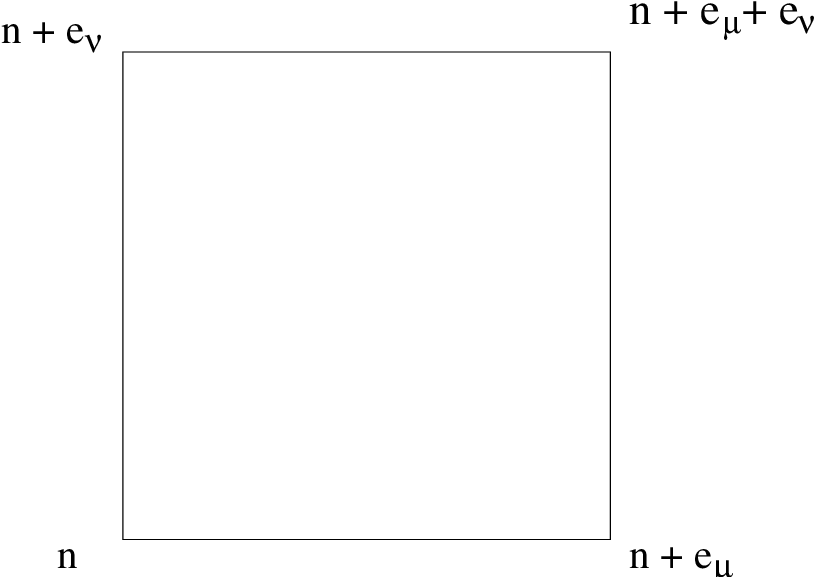}                  
     \ec
\caption{An elementary plaquette.}
 \label{plaquette}
    \end{figure}     

Thus, we introduce a  four-dimensional hypercubic lattice. The nodes of 
 the lattice
 are labelled by integer 4-vectors $n_\alpha$. Let us define on each {\it link}
 of the lattice a unitary matrix $U_{n,n + e_\mu} = U^\dagger _{n+ e_\mu ,n} \in SU(N)$, where 
 $e_1 = (1,0,0,0)$, 
 etc. For each {\it plaquette} (a two-dimensional face) of the lattice 
 labelled by its corner $n_\alpha$ with all the components less or equal than
 the corresponding components  for other three corners, and by two directional vectors
 $e_\mu$, $e_\nu$ (see Fig.~\ref{plaquette}), we define \cite{Wilson}
  \be
  \label{Wplaq}
W_{n, \mu\nu} \ =\ \frac 1N {\rm Re \ Tr} \left\{ 
 U_{n, n + e_\nu} U_{n + e_\nu, n + e_\mu + e_\nu}
 U_{n + e_\mu + e_\nu, n + e_\mu} U_{n + e_\mu, n} 
 \right\}.
  \ee
 (For $SU(2)$,  we need not take the real part  in Eq. (\ref{Wplaq}) as the trace of any $SU(2)$ matrix is real.)
Consider the integral 
 \be
 \label{pathlat}
 Z_{\rm lat} \ = \int \prod_{\rm links} {\cal D}U_{\rm link} \exp \left\{ - \frac {2N}{g^2}
 \sum_{\rm plaq} (1 - W_{\rm plaq}) \right\},
 \ee
where ${\cal D}U$ is the Haar measure
\index{Haar measure} on the group.
We are going to show that the exponent in Eq.~(\ref{pathlat})  is a
correct discrete approximation to the Euclidean
action of  continuum Yang--Mills theory and hence the integral 
(\ref{pathlat}) is a reasonable discrete approximation of the path integral,
 \be
  \label{ZYM}
  Z_{\rm cont} \ = \int \prod_{\mbox{\boldmath\scriptsize $x$}, \tau}  dA_\mu^a(\vec{x}, \tau)
  \exp \left\{ - \frac 1{2g^2}  \int d^4x \,  {\rm Tr}  \{\hat F_{\mu\nu} \hat F_{\mu\nu} \} \right\}. 
\ee
We assume  the actual spacing of our Euclidean lattice $a$ to be small compared
 to the characteristic scale $\Lambda_{\rm QCD}$ of  the theory.
 Let us associate $U_{n, n  + e_\mu}$ with the parallel transporter 
$ \exp\{-i a  \hat A_\mu e_\mu\}$ along the corresponding link where the $ \hat A_\mu$ may be defined at the point $n$. Then the quantity under the trace in Eq.~(\ref{Wplaq}) 
 is nothing but the {\it Wilson loop} on the closed contour $C$ around the  plaquette:\footnote{The symbol $P$ means a path-ordered product.}
\be
\lb{Wmncont}
 W_{n, \mu\nu} =\ \frac 1N {\rm Re Tr} \left[ P \exp \left\{- i \oint_C \hat A_\mu(\xi) \xi_\mu \right\} \right]  .
\ee
For small contours, this reduces to 
\be 
\lb{Wmnlim}
 W_{n, \mu\nu} = \ 1 - \frac {a^4}{2N} {\rm Tr} \{\hat F_{\mu\nu} \hat F_{\mu\nu}\} + O(a^6).
\ee 
Thus, we see that, indeed,
   $$ S_{\rm lat} \to  \frac{2N}{g^2} a^4 \sum_n \frac 12 \sum_{\mu\nu}
\frac 1{2N} {\rm Tr} 
\{ \hat F_{\mu\nu}^2\}  + O(a^6) \to \int d^4x \ \frac 1{2g^2} {\rm Tr} 
 \{ \hat F_{\mu\nu}^2 \}$$
in the continuum limit (the identity $\sum_{\rm plaq} = 
\sum_n \sum_{\mu > \nu} =
\frac 12 \sum_n \sum_{\mu\nu}$ was 
used). 

The integral (\ref{pathlat}) is invariant under transformations 
\be
 \label{gaugelat}
 U_{n, n+ e_\mu} \ \to \ \Omega_{n} U_{n, n  + e_\mu} \Omega^\dagger_{n + e_\mu} ,
 \ee
where $\{\Omega_n\}$ is a set of unitary matrices defined in the nodes of 
 the lattice.  This is the lattice version of gauge invariance \p{gaugefin} of continuum theory. 

 The integral in (\ref{pathlat}) involves a discrete but still infinite number
 of variables. To make it finite, our lattice should have  finite size
 in both  spatial and  Euclidean time directions. In  practice, it is
  convenient to implement this by imposing periodic boundary conditions on the 
matrix  link variables:
  \be
  \label{perU}
  U_{n + L_\alpha e_\alpha, n  + e_\mu + L_\alpha e_\alpha}\ = \ U_{n,  n+ e_\mu} ,
 \ee
where the set of integers $L_\alpha = \{L_x, L_y, L_z, L_\tau\}$ characterizes 
the size of the  lattice
(the number of nodes in the corresponding direction); no summation
 over $\alpha$ is assumed. With the conditions 
(\ref{perU}),  the
 theory is effectively defined on a discrete four-dimensional torus.

Toroidal boundary 
 conditions are easier to handle than boundary conditions with rigid walls: 
 finite size effects, which are present in all practical numerical 
calculations are  less prominent for the torus. For these boundary effects not
to be  important, the physical
 length  of the torus $aL_\mu$ should be larger than the characteristic scale
 of the theory, while $a$ should be kept much smaller. In practical 
 calculations, ``much'' usually means at most 4--5 times. 
 Calculating the
 path integral (\ref{pathlat}) numerically on an asymmetric lattice with
 $L_x = L_y = L_z \gg L_\tau$  in the regime where  boundary effects due
to  finite Euclidean time extension are important, while  effects due to a finite spatial  size are not, one finds the  partition function  of the system at finite temperature $T = 1/(aL_\tau)$. The properties
 of the vacuum wave functional are explored in calculations on large symmetric
 lattices.

{\bf Problem 3}. Derive \p{Wmnlim}.

{\bf Solution}. Consider a plaquette in the plane $(\mu \nu)$. We have
\be
&& \!\!\!\!\!\!\!\!\!\!\!P \exp \left\{- i \oint_{\Box_{\mu\nu}} \hat A_\rho(\xi) \xi_\rho \right\} \ \approx \ \left[1 - ia \hat A_\mu(n)  - \frac {a^2}2 \hat A_\mu(n)^2 \right] \times \nn
 &&\!\!\!\!\!\!\!\!\!\!\! \left[1 - ia \hat A_\nu(n+ e_\mu)  - \frac {a^2}2 \hat A_\nu(n+e_\mu)^2 \right]
\left[1 + ia \hat A_\mu(n+e_\mu+e_\nu)  - \frac {a^2}2 \hat A_\mu(n+e_\mu+ e_\nu)^2\right] \times \nn
&&\!\!\!\!\!\!\!\!\!\!\!
\left[1 + ia \hat A_\nu(n+e_\nu)  - \frac {a^2}2 \hat A_\nu(n+e_\nu)^2 \right] = \nn
&&\!\!\!\!\!\!\!\!\!\!\!1 - ia^2 (\pd_\mu \hat A_\nu - \pd_\nu \hat A_\mu)(n) - a^2[\hat A_\mu, \hat A_\nu](n) + O(a^3) = 1 - ia^2 \hat F_{\mu\nu}(n) + O(a^3).
\ee
Plugging that in \p{Wmncont}, we arrive at \p{Wmnlim}.

\subsection{Including quarks: first try}

QCD involves quarks and, to define  the path integral in QCD, we need to
 handle fermionic fields on the lattice. Let us define to this end Grassmann
 variables $ \bar \psi_n, \ \psi_n$ in the nodes of the lattice for each
 quark flavor (color and Lorentz indices are not displayed). As a first 
and natural guess, let us write
the extra term in the action as follows,
  \be
  \label{fermlat}
  S^{\rm ferm. lat.} \ =\  -\frac{ia^3}2 \sum_{n, \mu}
  [ \bar\psi_n U_{n, n+e_\mu} \gamma_\mu^E \psi_{n + e_\mu}
  -  \bar \psi_n U_{n, n-e_\mu} \gamma_\mu^E \psi_{n - e_\mu}] \nonumber \\
 +\  ma^4 \sum_n
  \bar \psi_n \psi_n. \ \ \ \ 
 \ee
 This expression is invariant under gauge transformations \p{gaugelat} complemented with $\psi_n \to \Omega_n \psi_n$. We see that the action (\ref{fermlat}) reproduces the action
   \be
\label{SferE} 
S_{\rm ferm}^E \ =\ \int d^4x [- i \bar \psi \gamma_\mu^E ( \partial_\mu - 
i  \hat A_\mu ) \psi  + m  \bar \psi \psi]
   \ee
 in the continuum limit. Indeed, for  free fermions
 $$ - \frac{ia^3}2 \sum_{n, \mu}
   \bar \psi_n  \gamma_\mu^E [\psi_{n + e_\mu}
  -\psi_{n - e_\mu}] \ \to \ - i \int d^4x \,
  \bar \psi \gamma_\mu^E  \partial_\mu  \psi.$$
  Expanding  $U_{n , n + e_\mu} \equiv 1 - i a  \hat A_\mu 
+ O(a^2)$, we also restore  the
interaction term, and the last term in Eq.~(\ref{fermlat}) turns into the
continuum mass term.
  
Equation (\ref{fermlat})\ is called the ``na\"{\i}ve lattice fermion action'',
 and I have to
say that if the reader was convinced by the above reasoning  that, in the
continuum limit, it goes over to Eq.~(\ref{SferE}),  s/he was na\"{\i}ve, 
too. Our implicit assumption was that the fermion fields $\psi_n$ depend on
the lattice node $n$ in a smooth manner, so that the finite difference
$\psi_{n + e_\mu} \!- \psi_{n - e_\mu}$ goes over to the continuum derivative.
It turns out, however, that  fermion field configurations which behave as
$\psi_n \sim (-1)^{n_1}$ or  $\psi_n \sim (-1)^{n_2 + n_4}$ and change 
significantly at the microscopic lattice  scale  are equally
important for the lattice path integral. After carefully performing continuum limit, these wildly 
oscillating modes give rise to 15 extra light fermion species with the same 
mass, the so-called {\it doublers}.

To understand it, consider first free massless fermions. Let 
 \be
\label{derpernaz}
(\partial_\mu^+ \psi)_n \ =\ \frac 1a [ \psi_{n + e_\mu} - \psi_n ],\ \ \ 
(\partial_\mu^- \psi)_n \ =\ \frac 1a [ \psi_{n} - \psi_{n-e_\mu} ]
   \ee
be the forward and backward lattice derivative operators. The na\"{\i}ve lattice counterpart of the free 
massless Dirac operator, which enters the path integral \p{detDir}, 
 is\footnote{ In the continuum limit, ${\mathfrak D}_{\rm free}^0 $ goes over to $-i \gamma^E_\mu \pd_\mu$. It is anti-Hermitian,  and its eigenvalues are purely imaginary. The full continuum action of a massive Euclidean fermion in gauge background was written in \p{SferE}:
\be
S^E_{\rm ferm} \ =\ \int d^4 x \, \bar \psi (m - i /\!\!\!\!{\cal D}) \psi .
\ee
}
   \be
  \label{Dirlatnai}
{\mathfrak D}_{\rm free}^0 \ =\  -\frac i2 \gamma^E_\mu (\partial_\mu^+ 
+ \partial_\mu^-) .
  \ee
The eigenfunctions of ${\mathfrak D}_{\rm free}^0$ are characterized by the
Euclidean 4-momentum $p_\mu$,
\be \
\psi_n^{(p)} \ =\ C_p e^{ia  p_\mu n_\mu } \ ,
\ee
where $C_p$ is a constant Grassmann bispinor. The eigenvalue equation
${\mathfrak D}_{\rm free}^0  \psi^{(p)}_n =  -i\lambda_p  \psi^{(p)}_n $ implies
 \be
\label{momlateig}
\left[ \frac 1a \gamma_\mu \sin (a p_\mu) \right] C_p \ =\  -i \lambda_p C_p
 \ee
with 
 \be
 \label{latfrspec}
\lambda_p \ =\ \pm \frac 1a \sqrt{\sum_\mu \sin^2 (a p_\mu)} .
  \ee

When $ap_\mu \ll 1$, we reproduce the continuum massless fermions with the
spectrum $\lambda_p = \pm \sqrt{p_\mu^2}$. Each eigenvalue 
(\ref{latfrspec}) is doubly degenerate due to two possible polarizations. The 
eigenfunctions with negative $\lambda_p$ are obtained from the ones 
with positive
$\lambda_p$ by multiplication by $\gamma^5$. Indeed, ${\mathfrak D}_{\rm free}^0$ anticommutes with $\gamma^5$. 
Let ${\mathfrak D}_{\rm free}^0 u = \lambda u$. Then ${\mathfrak D}_{\rm free}^0 (\gamma^5 u) = - \gamma^5 {\mathfrak D}_{\rm free}^0 u = -\lambda \gamma^5 u$.

Note, however, that the lattice Dirac equation (\ref{momlateig}) has an
{\it additional} discrete symmetry\footnote{See {\bf Problem 4} at the end of this section. One can observe that  $\hat Q_\mu \bar \psi_n$ and $\hat Q_\mu \psi_n$ stay mutually conjugated if $\bar \psi_n$ and $\psi_n$ were. This feature is not in fact  necessary, bearing in mind that $\bar\psi_n$ and $\psi_n$ should be treated as independent variables in Euclidean formulation, but it is still convenient.}
 $(Z_2)^4$:
\be
\lb{Qmu}
\hat Q_\mu \psi_n \ =\ (-1)^{n_\mu} \gamma^E_\mu \gamma^5 \psi_n, \qquad \hat Q_\mu \bar \psi_n \ =\ (-1)^{n_\mu +1} \bar \psi_n  \gamma^5 \gamma^E_\mu. 
 \ee 
For any eigenfunction 
$\psi_n^{(p)}$, the function 
$ \hat Q_\mu \psi_n^{(p)} $
 is also an eigenfunction of 
${\mathfrak D}_{\rm free}^0$ with the same eigenvalue $\lambda_p$. The operators $\hat Q_\mu$ commute with 
${\mathfrak D}_{\rm free}^0$ and anticommute with each other:
$$ \hat Q_\mu \hat Q_\mu + \hat Q_\mu \hat Q_\mu \ =\ 2\delta_{\mu\nu}.$$
The functions 
\be
\!\!\!\!\!\!\!\!\!\! \psi_n^{(p)}, \quad \hat Q_\mu \psi_n^{(p)}, \quad \hat Q_{[\mu} \hat Q_{\nu]} \psi_n^{(p)}, \quad 
\hat Q_{[\mu} \hat Q_{\nu} Q_{\lambda]} \psi_n^{(p)}, \quad \hat Q_{[\mu} \hat Q_{\nu} Q_{\lambda} Q_{\rho]} \psi_n^{(p)}
\ee
form a degenerate 16-plet. 

 In the free
case, each eigenstate of the na\"{\i}ve Dirac operator
 is not just
16-fold, but 32-fold degenerate due to polarizations. In the interacting case
(on a generic gauge field background), polarization is not 
a good quantum number, but the 16-fold degeneracy of all eigenstates of ${\mathfrak D}$ still
holds. The na\"{\i}ve lattice
Dirac operator  in Eq.~(\ref{fermlat}), which can be written in the form 
${\mathfrak D}^0 = - \frac i2 \gamma^E_\mu ({\cal D}_\mu^+ +
{\cal D}_\mu^-) $, where
\be
\label{dercovlat}
({\cal D}_\mu^+ \psi)_n &=&\frac 1a \left(\psi_{n + e_\mu}
U_{n, n+e_\mu} - \psi_n \right)\ ,\nonumber\\
({\cal D}_\mu^- \psi)_n &=&\frac 1a \left(\psi_n  - 
\psi_{n - e_\mu} U_{n, n-e_\mu}  \right) 
   \ee
are the covariant lattice forward and backward derivatives,
still enjoys the symmetries \p{Qmu}
 
We see that if an eigenstate $\psi_n$ changes smoothly 
from node to node, its 15 doublers
 wildly oscillate on the microscopic lattice spacing scale.
We might call these modes ``unphysical'', but they would not listen to us and 
contaminate with a vengeance any numerical lattice calculation we might wish
to do. Some way to get rid of them should be suggested, otherwise QCD, the
theory involving only 6 quarks with different masses, would not be 
operationally defined.

Simple-minded modifications of this na\"{\i}ve action, which leave only one light 
fermion for each flavor in the continuum limit, do not respect the  
chiral invariance  of the theory. It is a serious problem, and its solution is not easy.  At the end of the day,  we will see, however, that all these difficulties can be overcome
and a good consistent definition of the QCD path integral  exists.

 The action (\ref{fermlat}), as well as its 
sophistications to be discussed soon, are  bilinear 
in $ \bar \psi_n,
  \ \psi_n$. The fermionic part of the path integral has the form
  \be
  \label{detDir}
   \int \prod_{i} d \bar \psi_i d\psi_{i} \exp \{- {\mathfrak D}_{ij}
   \bar \psi_i \psi_j \} 
   \ , \ee
where $i \equiv (n, \alpha)$\,, with $\alpha$ marking both the color
and Lorentz spinor index, and ${\mathfrak D}_{ij}$ is a matrix turning 
into the
Euclidean Dirac operator $i \,/\!\!\!\!{\cal D}^E - m$ in the continuum
limit. The Grassmann integral gives the determinant,  $\det \|{\mathfrak D} \|$. Thus, the full path integral for
QCD has the form
  \be
  \label{ZQCD}
Z = \int \prod_{\rm links} {\cal D}U_{\rm link} \prod_f \det \|{\mathfrak D}_f \|
 \exp \left\{ - \frac {2N}{g^2}
 \sum_{\rm plaq} (1 - W_{\rm plaq}) \right\}
   \ , \ee
where $\prod_f$ runs over all quark flavours.

Numerical calculations of the
integrals like (\ref{ZQCD})  are technically very difficult: not only one
has to do a multidimensional integral, but also the {\it integrand} becomes
very complicated involving the determinant of a large matrix. With modern computers and clever algorithms, such calculations are still possible.

\vspace{.2cm}

{\bf Problem 4}. Verify that ${\mathfrak D}^0$ is invariant under the transformations \p{Qmu}.

\vspace{.1cm}

{\bf Solution}. The na\"{\i}ve lattice Euclidean fermion action reads
\be
S^E_{\rm ferm} \ =\ \frac 1{2a} \sum_{n\mu} \bar \psi_n \gamma^E_\mu (U_{n,n+e_\mu} \psi_{n + e_\mu} - 
U_{n,n-e_\mu} \psi_{n - e_\mu}) .
\ee
Consider the term corresponding to a particular node $n = (0,0,0,0)$ of the lattice  and look how the variables $\psi_n$ (call it $\psi_0$) and $\psi_{n + e_\mu}$ (call them $\psi_\mu$) transform under the action of $\hat Q_1$. The definition \p{Qmu} gives:
\be 
&&\hat Q_1 \psi_{0,2,3,4}   = \gamma^E_1 \gamma^5 \psi_{0,2,3,4},  \qquad  \quad \hat Q_1 \bar \psi_{0,2,3,4}   = - \bar \psi_{0,2,3,4}  \gamma^5  \gamma^E_1 \nn
 &&\hat Q_1 \psi_1 =  -  \gamma^E_1 \gamma^5  \psi_1, \qquad  \qquad  \qquad \hat Q_1 \bar\psi_1 =  \bar\psi_1   \gamma^5 \gamma^E_1  \nonumber .
\ee
Then $\bar \psi_0 \gamma_1 \psi_1$ transforms as
$$ \bar \psi_0 \gamma^E_1 \psi_1 \ \to\ (- \bar \psi_0 \gamma^5 \gamma^E_1) \gamma^E_1 ( -  \gamma^E_1 \gamma^5  \psi_1) = 
-  \bar \psi_0 \gamma^5 \gamma^E_1 \gamma^5 \psi_1 =   \bar \psi_0 \gamma^E_1 \psi_1,
$$
i.e. it does {\it not} transform
[the property \p{Eucl-com} was used]. Similarly for the other terms:
$$
\bar \psi_0 \gamma^E_2 \psi_2 \ \to\ (- \bar \psi_0 \gamma^5 \gamma^E_1) \gamma^E_2 (   \gamma^E_1 \gamma^5  \psi_2) = 
-  \bar \psi_0 \gamma^5 \gamma^E_2 \gamma^5 \psi_2 =   \bar \psi_0 \gamma^E_2 \psi_2
$$
and the same for $\bar \psi_0 \gamma^E_3 \psi_3$ and $\bar \psi_0 \gamma^E_4 \psi_4$.

\vspace{.2cm}

\subsection{Nielsen--Ninomiya's No-go theorem}

The problem to construct a viable lattice counterpart of the continuum Dirac operator  is not at all simple. Let us look for some other lattice Dirac operator ${\mathfrak D} \neq {\mathfrak D}^0$ satisfying the following
four natural conditions:
 \begin{enumerate}
\item At distances much larger than the lattice spacing $a$, 
${\mathfrak D} \to - i\,/\!\!\!\!{\cal D}^E$, giving rise to a massless fermion in
the continuum limit.\footnote{Adding a finite mass term to ${\mathfrak D}$ 
provides no difficulties [see Eq.~(\ref{fermlat})].}

\item All the modes of ${\mathfrak D}$ not associated with the latter are of 
order $1/a$ (no doublers!).

\item ${\mathfrak D}$ is local. In other words, the matrix elements
${\mathfrak D}_{n n'} $ decay exponentially fast at large distances
$|n - n'| \gg 1$. 

\item Chiral symmetries (\ref{U1chir}), (\ref{symax}) of the massless 
fermionic action are 
not broken
by the regularization  explicitly. [The singlet axial symmetry  (\ref{U1chir})
 is eventually going to be
broken due to noninvariance of the  fermionic  measure, but we require
the absence of {\it explicit} breaking in the regularized action.]  This seems 
to imply the condition ${\mathfrak D}\gamma^5 +
\gamma^5 {\mathfrak D} = 0$.
 
\end{enumerate}

\begin{figure}        
\bc
 \includegraphics[width=.8\textwidth]{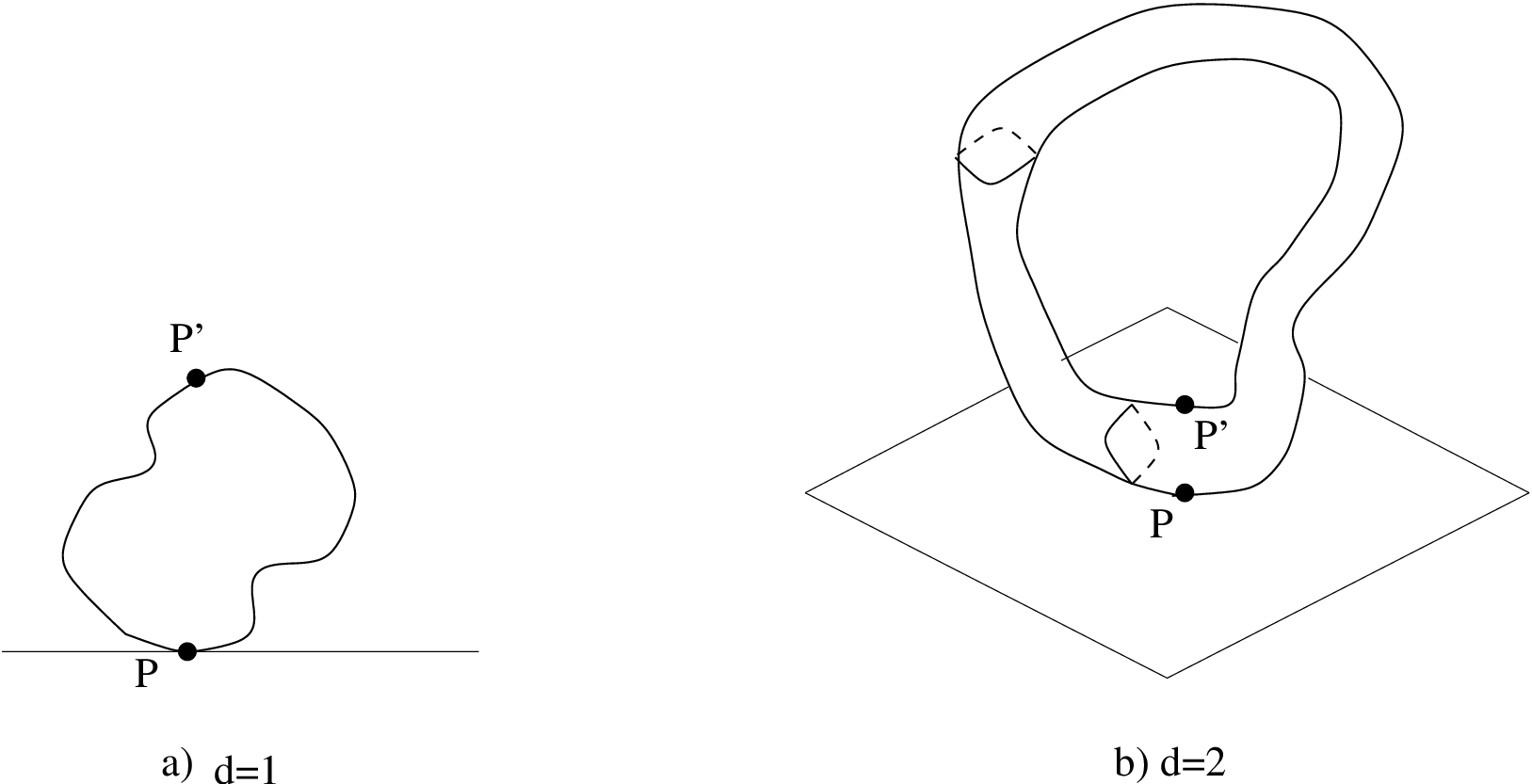}
\ec        
\caption{Nielsen--Ninomiya theorem. $P$ and $P'$ are the 
             different zeros of the lattice Dirac operator.}
\label{proobr}
\end{figure}

The no-go theorem due to  Nielsen and Ninomiya \cite{no go} tells us, however, that such
a ${\mathfrak D}$ {\it does not exist}. To understand it, consider first 
the free fermion case. The momentum $p_\mu$ is then a good quantum number, and
the Dirac operator in the momentum representation has the form
 \be
{\mathfrak D} (p) \ =\ \gamma^E_\mu F_\mu (p) + G(p).
 \ee
Condition 4 tells us that $G(p) = 0$. Condition 1 implies
that $F_\mu(p) = p_\mu + O(ap^2) $ for $ap_\mu \ll 1$. Now,
$F_\mu (p)$ are four periodic functions of their four arguments $p_\mu$ with 
period $2\pi/a$. They thus realize  a smooth map $T^4 \to R^4$. A look at Fig.~\ref{proobr}
can convince the reader that the point $F_\mu = 0$ on $R^4$ where our torus touches
it always has at least one more pre-image. Their intuition would not
betray them: this statement can be proven in a rigorous mathematical manner. 
Basically, it follows from the fact that the degree of a map
  $T^d \to R^d$ is zero, which means that \cite{DNF}
 \be
\label{degrmap}
\sum_{
\begin{array}{c}
\mbox{{\small   pre-images}}\\  \mbox{{\small  of} $P$ }
\end{array}
}\!\!\!{\rm sign}\ 
\left[ \det \|
\partial_\nu F_\mu(p) \| \right] \ =\ 0.
  \ee
 As the Jacobian of the mapping $p_\mu \to F_\mu(p)$ 
is equal to 1 at the point $p_\mu = 0$,
Eq.~(\ref{degrmap}) implies that some other pre-images of zero, i.e.\ some other
solutions of the equation system $F_\mu (p) = 0$ should be present (one can
have just one extra solution as in Fig.~\ref{proobr} or more: for the ``round
upright torus'' $F_\mu(p) = \sin(ap_\mu)/a$, there are $2^d - 1$ extra
solutions). And that means the presence of doublers
in contradiction with the condition 2.

The only remaining possibility is that $F_\mu (p)$ are not continuous. Besides
being ugly, it also contradicts   condition 3: the matrix elements
${\mathfrak D}_{n n'} = {\mathfrak D}(n - n') = \gamma^E_\mu \tilde F_\mu (n - n')$
are actually the Fourier coefficients of the periodic function
$\gamma^E_\mu F_\mu (p)$. If the latter is discontinuous, the Fourier
coefficients cannot decay faster than $1/|n - n'|$ (otherwise, the Fourier
series would converge uniformly on the torus $p_\mu \in \left[0, \frac{2\pi}a 
\right)$ and the 
sum of such a series would be continuous). 

We have proven that a lattice Dirac operator satisfying the above four
conditions  cannot be found for free fermions, but this  also means that it 
cannot be 
found for QCD: any ${\mathfrak D}$ having this property should satisfy our list   
 for any smooth set of link variables 
and, in particular, for the set $U_{\rm each\ link} = 1$ corresponding to free theory.

\subsection{Ways to go $\scriptstyle{\Box}$ The Ginsparg--Wilson way.}

If we still want to build up a lattice version of QCD, we have to relax
at least one of our four conditions. Conditions 1 and 2 are, however, 
indispensible: a lattice theory where they do not hold just has nothing to
 do with QCD. Therefore, either  locality  or manifest chiral invariance of the lattice
action should be abandoned.

One of the possible procedures is that only {\it one mode} of each degenerate
16-plet of ${\mathfrak D}^0$ is taken into account in the fermionic determinant
and in the spectral decomposition of fermion Green's functions,
 \be
 \label{speclat}
\langle \psi_n \bar \psi_{n'} \rangle \ =\ \sum_{k}
\frac {u_k (n) u_k^\dagger (n') }{m - i \lambda_k},
 \ee 
 where $u_k (n)$ describe the $k$-th eigenmode of  ${\mathfrak D}^0$ as a 
function of the node.  This amounts to choosing the lattice Dirac operator in 
the form
$({\mathfrak D}^0)^{1/16}$,  which is not local. A similar method is sometimes
used in  practical lattice calculations, but besides purely technical
inconveniences  it is 
unsatisfactory from a philosophical viewpoint: we {\it would} 
like to have a local lattice approximation for a local field theory.  

But then the chiral invariance like (\ref{U1chir}) is  necessarily lost. 
Though renouncing 
chiral invariance is also not desirable --- when regularizing the theory, we
should try to preserve as much of its symmetries as possible ---  it can still be
considered  the least of evils. After all, lattice regularization also does not preserve other important symmetries of the continuum action --- translational and Lorentz invariances. They are restored only in the limit $a \to 0$. One can allow the same  for the chiral invariance.

Two ways of chiral noninvariant lattice regularization
have been known for some time 
and used in practical calculations: {\it (i) Wilson}
 fermions \cite{Wilson} and
{\it (ii) Kogut--Susskind}
 or {\it staggered} fermions \cite{Kogut}. 
We will describe
here the first method which consists in adding to ${\mathfrak D}^0$ the term 
$ - \frac {ra}2 {\cal D}_\mu^+  {\cal D}_\mu^- = - \frac {ra}2  {\cal D}_\mu^- 
 {\cal D}_\mu^+$  with the covariant lattice derivatives defined in \p{dercovlat}.
 Thus, the Wilson--Dirac
 operator is defined as \cite{Gatt}
 \be
 \label{WilsDir}
{\mathfrak D}^W \ =\ - \frac i2 \gamma^E_\mu \left( {\cal D}_\mu^+ + 
{\cal D}_\mu^- \right) - \frac {ra}2 {\cal D}_\mu^+  {\cal D}_\mu^- 
 \ee
with
\be
\!\!\!({\cal D}_\mu^+  {\cal D}_\mu^- \psi)_n =  \frac 1{a^2} \sum_{\mu =1}^4 \left( U_{n,n+e_\mu} \psi_{n+e_\mu} + 
 U_{n,n-e_\mu} \psi_{n-e_\mu}  - 2 \psi_n   \right).
\ee 
Here $r$ is an arbitrary nonzero positive constant. 

The first term in Eq. (\ref{WilsDir}) (call it  $iB$) is anti-Hermitian
while the second term (call it $A$) is Hermitian. 
The property
\be
\lb{obklad}
\gamma^5 {\mathfrak D}^W  \gamma^5 \ =\ ({\mathfrak D}^W )^\dagger
\ee
holds.
$A$ and $B$ do not 
commute for a generic gauge field configuration and cannot be simultaneously
diagonalized. 
A generic matrix ${\mathfrak D}^W = A + iB$ still can be diagonalized, but 
the corresponding
transformation matrix is not  unitary and  the eigenvectors 
of ${\mathfrak D}^W$ are not orthogonal to each other. 
This might be  not so nice, but it does  not prevent one  to 
determine the spectrum of ${\mathfrak D}^W$ (the roots of the characteristic
equation) and compute  the
determinant of ${\mathfrak D}^W +  m$ which enters the  lattice approximation
of the partition function of QCD.

The essential properties of ${\mathfrak D}^W$ can be understood
by studying the case of free fermions. The situation is much simpler
here than in the interacting case. The Hermitian and anti-Hermitian
parts of ${\mathfrak D}^W_{\rm free}$ can now be simultaneously diagonalized,
and this can be done explicitly. We have
$ {\cal D}_\mu^\pm \to  \partial_\mu^\pm$ and
$$ ( \partial_\mu^+  \partial_\mu^- \psi)_n = \frac 1{a^2} \sum_\mu
\left( \psi_{n + e_\mu} +  \psi_{n - e_\mu} - 2 \psi_n \right)\ , $$
which is the lattice Laplacian. Passing to the momentum representation, we obtain
 \be
\label{DWmom}
{\mathfrak D}^W_{\rm free} (p) \ =\ \frac 1a \gamma^E_\mu \sin (ap_\mu)
+ \frac {2r}a \sum_\mu \sin^2 \left( \frac {ap_\mu}2 \right)  .
 \ee
The second term has the form of a momentum-dependent mass. For small $p_\mu \ll
1/a$, it can be neglected and the continuum massless Dirac operator is
reproduced. In contrast to ${\mathfrak D}^0$, the operator (\ref{DWmom}) 
is not anti-Hermitian, and its eigenvalues are complex. What is important is
that, for  $p_\mu$ that are not small, the absolute values of the eigenvalues of 
 ${\mathfrak D}^W_{\rm free}$\,, 
    \be
    \label{eigWil}
 -i \lambda_p^W \ =\ \pm \frac ia  \sqrt{\sum_\mu \sin^2 (a p_\mu)}
  + \frac {2r}a \sum_\mu \sin^2 \left( \frac {ap_\mu}2 \right) , 
   \ee
are of order $1/a$. The doublers
disappear. At 
$p_\mu = \left( \frac \pi a, 0,0,0 \right)$, the eigenvalue 
(\ref{eigWil}) is $\frac {2r}a$; at 
$p_\mu = \left( 0, \frac \pi a, 0, \frac \pi a
 \right)$, it is $\frac {4r}a$, etc. 

The chiral symmetry is broken, however, and it is messy. In principle, when
the continuum limit $a \to 0$ is taken, the effects due to the breaking
of $\gamma^5$ invariance must be suppressed, but in this particular 
problem the continuum limit with restoration of chiral symmetry is rather
slow to reach, and, though  most experts think that it is
eventually reached, this has  not been shown  quite rigorously. 
In particular, it is difficult to make the pions light.
In practical calculations this is achieved by introducing 
a large bare quark mass of order $g^2(a)/a$ and fine-tuning it so that the 
effects due to
two chiral noninvariant terms --- the Wilson term and the bare quark term
 --- would cancel each other. Needless to say, this is a rather artificial
and unaesthetic procedure.

As we see, this Nielsen--Ninomiya puzzle represented a complicated logical knot that resisted attempts to untie 
it. A remarkable observation \cite{Neuberger, Luscher} was that the best strategy here was 
to follow the example of the Alexander the Great and just cut  it through!
The adequate sword was forged back in 1982 by Ginsparg
 and Wilson \cite{Ginsparg}. They
suggested to consider the lattice Dirac operators satisfying  the relation
 \be
 \label{GWrel}
\gamma^5 {\mathfrak D} + {\mathfrak D} \gamma^5 \ =\ a {\mathfrak D} \gamma^5
{\mathfrak D}
  . \ee
The anticommutator $\{ {\mathfrak D}, \gamma^5 \}$ does not vanish which means
that the lattice action is not invariant with respect to the chiral 
transformations
 \be
 \label{chirnai}
\delta \psi_n =  -i\alpha \gamma^5 \psi_n\ ,\ \ \  \ \delta \bar \psi_n = 
-i\alpha \bar \psi_n  \gamma^5 ,
 \ee
a lattice Euclidean counterpart of Eq.~(\ref{U1chir}).

It took 16 years to realize that the lattice
 fermion action
 \be
 \label{SFgen}
S_F \ =\ a^4 \sum_{n n'} \bar \psi_n {\mathfrak D}_{n n'} \psi_{n'}
\ee
(color and spinor indices being suppressed), with ${\mathfrak D}$ 
satisfying the relation ({\ref{GWrel}), {\it is} invariant with respect to the
following transformations:\footnote{In this case, $\delta \bar \psi \neq (\delta \psi)^\dagger$ not only because of the presence of $i$ in the right-hand sides but also due to the fact that ${\mathfrak D}$ is not Hermitian.}
 \be
 \label{transLu}
\delta \psi &=&-i \alpha \gamma^5 \left[1 - \frac 12 a{\mathfrak D} \right] \psi\ ,
 \nonumber \\
\delta \bar \psi &=&-i \alpha \bar \psi  
\left[1 - \frac 12 a{\mathfrak D} \right] \gamma^5 .
 \ee
If ${\mathfrak D}$ is local (in the sense of condition 3 in the 
Nielsen--Ninomiya list), Eq.~(\ref{transLu}) is as good a lattice approximation
of the continuous chiral symmetry (\ref{U1chir}) as the trivial 
(\ref{chirnai}).
In particular, the pions would automatically be light (massless in the chiral
limit), and no fine tuning is required. In addition,  condition 4 above is no longer  
satisfied  and one can now hope  to find a local ${\mathfrak D}$
not involving doublers.
 The problem is still not trivial: as we will see later (from the solution of  {\bf Problem 5}),  not all solutions of the Ginsparg--Wilson
 relation (\ref{GWrel}) eliminate the doublers.  The simplest  solution that {\it does} was suggested
by  Neuberger in Ref. \cite{Neuberger}.
It has the form\footnote{This construction is closely related to the so-called {\it overlap representation} of the Dirac operator \cite{Rajamani}. But we will not go into further details here.}
 \be
\label{Neuberger}
{\mathfrak D} \ =\ \frac 1a \left[ 1 - A 
\frac 1 {\sqrt{A^\dagger A}} \right],
 \ee
where $A = 1 - a {\mathfrak D}^W $ and ${\mathfrak D}^W$ 
is the Wilson--Dirac
 operator (\ref{WilsDir}) 
with $r > 1/2$. The Hermitian matrix $A^\dagger A$ has only positive eigenvalues. Out of many possible square roots, we choose one where the eigenvalues are also positive. 

The operator (\ref{Neuberger})
satisfies the Ginsparg--Wilson
 relation. To see that, represent $a{\mathfrak D} = 1 -V$. Note that the operator $V$ is unitary:
\be
&&V V^\dagger \ =\  A (A^\dagger A)^{-1/2} (A^\dagger A)^{-1/2} A^\dagger\ =\ A[A^{-1} (A^\dagger)^{-1}] A^\dagger \ =\ 
 \mathbb{1}; \nn
&&V^\dagger V = (A^\dagger A)^{-1/2} A^\dagger A (A^\dagger A)^{-1/2}\  = \  \mathbb{1}. \nonumber
\ee
Note also that\footnote{It is basically a collorary of \p{obklad}. It follows from \p{obklad} immediately for the free fermions when $A^\dagger A = A A^\dagger$, but this property also holds in the interacting case (see {\bf Problem 6}).}  \be
\lb{obklad1}
\gamma^5 V \gamma^5 = V^\dagger. 
 \ee   
Compare
$$ \gamma^5(1-V) + (1-V) \gamma^5 \qquad {\rm with} \qquad (1-V)\gamma^5 (1-V).
$$
The LHS and RHS coincide iff $\gamma^5 = V\gamma^5 V$ or $V\gamma^5 V \gamma^5 = \mathbb{1}$. But that is true due to \p{obklad1} and unitarity of $V$. 

It follows from \p{GWrel} that, in contrast to ${\mathfrak D}^W$, ${\mathfrak D}$ satisfies the property
$$ [{\mathfrak D}, {\mathfrak D}^\dagger] \  =\ {\mathfrak D} \gamma^5 {\mathfrak D} \gamma^5 -  \gamma^5 {\mathfrak D} \gamma^5  {\mathfrak D} \ =\ \frac 1a \left[
({\mathfrak D}\gamma^5 + \gamma^5 {\mathfrak D}) \gamma^5  - \gamma^5
({\mathfrak D}\gamma^5 + \gamma^5 {\mathfrak D})\right]  \ =\ 0,
$$
 allowing for a  
 simultaneous diagonalization of  its Hermitian and  anti-Hermitian part. 

In particular, for $r =1$ and for the
free fermions, we have 
 \be
\label{Neumom}
 a {\mathfrak D}(p) \ =\ 1 - \ \frac {1 - 2 \sum_\mu \sin^2 \left( \frac
{ap_\mu}2 \right) -  \sum_\mu \gamma_\mu \sin (ap_\mu)}
{\left[ 1 + 8 \sum_{\mu < \nu} \sin^2 \left( \frac {ap_\mu}2 \right)
 \sin^2 \left( \frac {ap_\nu}2 \right) \right]^{1/2}}.
 \ee
 The eigenvalues of (\ref{Neumom}) are different from zero provided 
$p_\mu \neq 0$. In particular, for former doublers, 
$$p_\mu =  \left( \frac \pi a, 0,0,0 
\right),  \ \  p_\mu =  \left( \frac \pi a, \frac \pi a ,0,0 \right), \ \  
p_\mu =  \left( \frac \pi a, \frac \pi a, \frac \pi a,0 \right)\ ,$$
 and
$$p_\mu =  \left( \frac \pi a, \frac \pi a, \frac \pi a ,  \frac \pi a 
\right),$$ 
the eigenvalues $-i\lambda$ of ${\mathfrak D}$ are all equal to $ 2/a$.
The doublers
 are absent. 

A second look at Eq. (\ref{Neumom}) reveals a 
beautiful feature displayed in Fig.~\ref{krugfig}\, : the eigenvalues of
${\mathfrak D}$ lie on the circle
 \be
 \label{krug}
\left( {\rm Re} \ \lambda \right)^2 +
\left( {\rm Im }\ \lambda - \frac 1a \right)^2 \ =\ \frac 1{a^2} .
 \ee
This property holds also in the interacting case. As was mentioned,
the operator $V$ is unitary, i.e.\ its eigenvalues lie
on the circle $\{ e^{i\phi} \}$. And the eigenvalues of ${\mathfrak D}
= (1-V)/a$ lie on the shifted circle in Fig.~\ref{krugfig}.

 \begin{figure}
\bc
 \includegraphics[width=.6\textwidth]{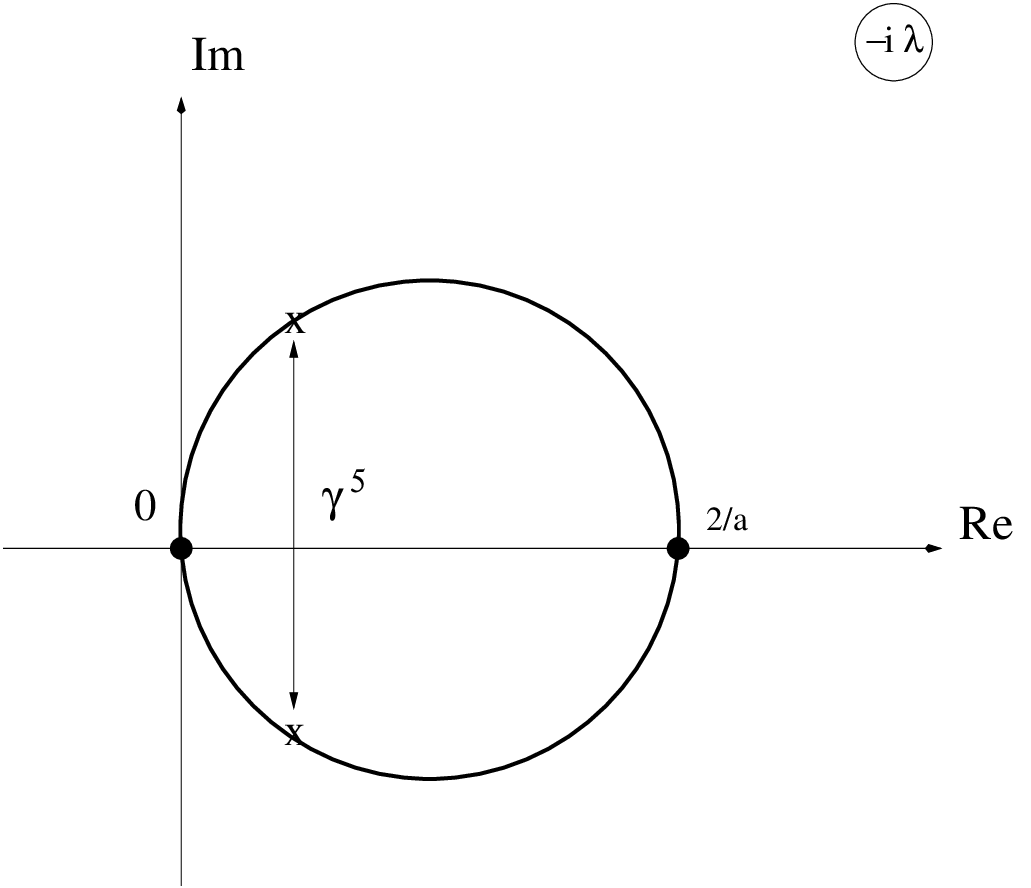}        
\ec
\caption{The circle of eigenvalues for  Neuberger's
operator. The
              eigenmodes with the eigenvalues marked by crosses are 
              related by a $\gamma^5$ transformation.}
\label{krugfig}
\end{figure}

The function (\ref{Neumom}) has singularities associated with the square
root, but they all occur at complex values of $p_\mu$. 
It is analytic on the torus $p_\mu \in \left[ 0, 
\frac {2\pi}a \right)$,  
which means that its Fourier image decays exponentially at large distances.
The Dirac operator thus constructed is local. In the interacting case, it
stays local if the gauge field is smooth enough, i.e. if the link variables 
$U_{n, n+e_\mu}$ are sufficiently close to 1. 

As was mentioned, the singlet axial symmetry (\ref{U1chir}) 
is anomalous, which
shows up in the noninvariance of the fermionic measure. The measure
$\prod_n d\bar \psi_n d \psi_n$ is obviously invariant, however, with respect
to the ultralocal transformations (\ref{chirnai}). This follows from the
fact that ${\rm Tr}\{ \gamma^5\} = 0$. On the other hand, Eq.~(\ref{lnJ}) 
relates
the modification of the measure to the operator trace of $\gamma^5$,
 \be
\label{optrace}
 {\rm \bf Tr}\{\gamma^5 \} \ =\ \int d^4x \sum_k u^\dagger_k(x) \gamma^5 
u_k(x), \ee
 which is thus also zero. For the na\"{\i}ve Dirac operator, only the zero
modes contribute to the sum (\ref{optrace}),  
which means that the number of the right-handed and the 
left-handed zero modes
 of ${\mathfrak D}^0$ should be equal.

This also follows from the fact that ${\mathfrak D}^0$ commutes with $\hat Q_\mu$ defined in Eq. (\ref{Qmu}). Any smooth eigenfunction $\psi_n^{(p)}$ of ${\mathfrak D}^0$  is accompanied by 15 doublers. It is not difficult to see that if  $\psi_n^{(p)}$ is, say, right-handed,  the eigenfunctions $ \hat Q_{[\mu} \hat Q_{\nu]} \psi_n^{(p)} $  and $\hat Q_{[\mu} \hat Q_\nu  \hat Q_{\lambda} \hat Q_{ \rho ]} \psi_n^{(p)} $ are  also right-handed, whereas $\hat Q_\mu \psi_n^{(p)}  $ and  $\hat Q_{[\mu} \hat Q_\nu \hat Q_{\lambda]} \psi_n^{(p)} $ are left-handed. 
In particular, instead of a single right-handed zero mode in (the lattice approximation for) the instanton background, we have a degenerate 16-plet with  8 right-handed and 8 left-handed modes which  are not   necessarily zero  modes  anymore. 

The vanishing of  the index of  ${\mathfrak D}^0$ is closely related to the identity  (\ref{degrmap}). Indeed, the sign of the Jacobian $\det \| \partial_\nu F_\mu(p) \|$ describes the orientation of the neighbourhood of a pre-image $P_i \in T^4$ with respect to the orientation of the tangent space $R^4$ into which it is mapped. This orientation is obviously related to chirality.

The absence of anomaly is one of the diseases of the na\"{\i}ve lattice Dirac 
operator.
 For the  operators of Ginsparg--Wilson
 type, the situation is 
different.
The chiral symmetry is now implemented as in Eq.~(\ref{transLu}) and,
generically, the measure is {\it not} invariant with respect to these 
transformations. We have, instead of Eq.~(\ref{lnJ}),
 \be
\label{lnJGW}
\ln J \ =\ -i\alpha \left[ n_R^{(0)} - n_L^{(0)} \right] \ =\ -i\alpha {\bf Tr} \left\{ \gamma^5 \left( 1 - \frac 12 a
{\mathfrak D} \right) \right\}.
 \ee
Even though  {\bf Tr}$\{ \gamma^5 \} = 0$,  {\bf Tr}$\{ \gamma^5  {\mathfrak D}\}$  
need not vanish and the anomaly
 is there. 

It is instructive to see how a nonzero operator trace on the
right side of Eq.~(\ref{lnJGW}) is realized in the basis 
including the eigenstates of ${\mathfrak D}$.
First, it is still true that, for any eigenfunction $u_k$ of ${\mathfrak D}$,
$\gamma^5 u_k$ is also an eigenfunction. However, in contrast to the  continuum or
na\"{\i}ve lattice Dirac
operators where the eigenvalues of $\gamma^5 u_k$ and $u_k$ were related as $\lambda_k' = -\lambda_k$, in the Ginsparg--Wilson
 case, the corresponding relation is 
$\lambda_k' \ =\ -\lambda_k^*$. Complex conjugation appears! Thus, for almost all eigenstates on the circle
in Fig.~\ref{krugfig}, $u_k$ and $\gamma^5 u_k$ have {\it different}
eigenvalues and are orthogonal to each other. Their contribution to the
trace vanishes. The only exception are two points on the circle, 
$\lambda = 0$ and $\lambda = 2i/a$, where $\lambda_k = \lambda_k'$ and the
eigenstates can have a definite chirality. But the ``heavy doublers" with
 $-i \lambda = \frac 2a$  obviously give zero contribution in Eq.~(\ref{lnJGW}).
Only zero modes
 of ${\mathfrak D}$ are relevant. We have derived the
{\it lattice index theorem} \cite{HLN}:
\be
 \label{indexlat}
n^0_R({\mathfrak D}) - n^0_L({\mathfrak D}) \ = \ 
- \frac 12 {\rm Tr} \{\gamma^5 a {\mathfrak D} \} .
 \ee 
The right side of Eq.~(\ref{indexlat}) is a functional depending of the link
variables $\{U\}$. By definition, it is given by a sum over all lattice nodes
of some complicated expression, which  goes
over to the topological charge (\ref{k4F}) in the continuum limit.\footnote{To the best of my knowledge, this expression has never been explicitly written out. It would be an interesting problem to do so.} 

\vspace{.4cm}

{\bf Problem 5}. Consider the free Wilson--Dirac operator
 (\ref{DWmom}) with 
arbitrary
$r$ and analyze the corresponding Neuberger's operator
  (\ref{Neuberger}).
Show that only the values $r > 1/2$ are admissible.

\vspace{.1cm}

{\bf Solution}. As was mentioned above, the eigenvalues of the free 
Wilson--Dirac
 operator (\ref{WilsDir}) 
which correspond to  the na\"{\i}ve doublers
 are $-i\lambda^W = 2rl/a$, where $l$
is the number of nonzero components $\pi/a$ of the momenta $p_\mu$. The 
corresponding eigenvalues of the operator  (\ref{Neuberger}) are 
 \be
\label{doubll}
 -i\lambda  \ =\ \frac 1a \left[ 1 - \frac{1 - 2rl}{|1 - 2rl|} \right] .
 \ee
We see that, if $r < 1/2$, four doublers
 with $l=1$ become massless again, 
and that is not what we want. 

\vspace{1mm}
For $r <1/8$, {\it all} the doublers stay massless. Note that the 
operator (\ref{Neuberger}) satisfies the Ginsparg--Wilson
 relation for any
$r$,  which means that its eigenvalues still lie on the circle in Fig.~\ref{krugfig}. But only a part of the circle is covered.

Massless doublers
 may appear also for $r$ slightly
exceeding $1/2$ if gauge fields are present.

\vspace{.2cm}

{\bf Problem 6}. Derive the property \p{obklad1}.

\vspace{.1cm}

{\bf Solution}.
Consider an analytic function $f(A^\dagger A)$ and its Taylor expansion $f(y) = a_0 + a_1 y + \cdots$. Then
\be
&&
\gamma^5 A f(A^\dagger A) \gamma^5 \ =  \gamma^5 A (a_0 + a_1 A^\dagger A
+ a_2 A^\dagger A A^\dagger A + \cdots) \gamma^5  \nn 
 && =\ a_0  A^\dagger + a_1 A^\dagger A A^\dagger + a_2 A^\dagger A A^\dagger A A^\dagger  + \cdots \nn 
&&= \ f(A^\dagger A) A^\dagger  = [A f(A^\dagger A)]^\dagger. 
\ee
The reader may complain that $f(y) = 1/\sqrt{y}$ is not analytic at $y=0$. True, but one can expand it in  Taylor series at any other point.  

\section{Aspects of chiral symmetry}
\setcounter{equation}0

We now abandon the lattice  and will  discuss in this section 
various aspects of chiral symmetry in the continuum limit. Sometimes, we
will think in terms of quarks, of the
 symmetries   (\ref{U1chir}) and  (\ref{symax}),
and of the order parameter (\ref{condfg}) associated with the spontaneous 
breaking of the flavor-nonsinglet symmetry. Sometimes, we will describe the
system in terms of the pseudo-Goldstone
 degrees of freedom and the effective
Lagrangians
 (\ref{Leffchir2}), (\ref{Leffchir4}), and  (\ref{Lchirmass}). 
Sometimes, we will confront the two languages using the philosophy of
the {\it quark--hadron duality}
 ---   that is how  many
results discussed in this section, a bunch of beautiful
{\it exact theorems of QCD}, will be obtained.

\subsection{QCD inequalities $\scriptstyle{\Box}$ Vafa--Witten theorem.}

As was discussed above, the octet of pseudoscalar mesons $(\pi, K, 
\eta)$ can be interpreted as  that of the pseudo-Goldstone
particles appearing due to the  spontaneous 
chiral symmetry
breaking according to the pattern (\ref{patbreak}) in
the massless limit. This is the reason why 
the pseudoscalar mesons are lighter than
those with other quantum numbers.
It is interesting that the latter statement can be formulated as an exact 
theorem of QCD without any reference to the (experimental!) fact that the
chiral symmetry is broken.

Consider a QCD-like theory with at least two quark flavors 
and assume that these quarks (call them 
 $u$ and $d$) have equal masses $m_u = m_d = m$.
Consider a set of Euclidean correlators,
 \be
C_\Gamma(x,y) \ =\ \langle J^{\bar u d}(x)  J^{\bar d u}(y) \rangle_{\rm vac} 
,
\label{setcorr}
 \ee 
where $ J^{\bar u d} (x)$ are flavor-changing bilinear quark currents,
$ J^{\bar u d} = \bar u \Gamma d$, with Hermitian 
$$\Gamma = 1, \,\,\gamma^5,\,\,
i\gamma^E_\mu, \,\,\gamma^E_\mu \gamma^5 \ {\rm and} \    i\sigma_{\mu\nu} = \frac i2(\gamma^E_\mu \gamma^E_\nu - \gamma^E_\nu \gamma^E_\mu)$$
 At large distances, 
the correlators (\ref{setcorr}) decay exponentially 
 \be
\label{corrasmes}
C_\Gamma(x,y) \ \propto \ \exp\{-M_\Gamma|x-y| \} ,
  \ee
where $M_\Gamma$ is the mass of the lowest meson state in the corresponding 
channel.\footnote{We assume here that the quarks are confined, 
otherwise the whole
discussion is pointless.}

On the other hand, the correlators 
(\ref{setcorr}) of the quark currents can be
expressed as
 \be
\label{TrGSGS}
C_\Gamma(x,y) \ =\ - Z^{-1} \int d\mu_A {\rm Tr} 
\left\{ \Gamma {\cal G}_A(x,y) \Gamma {\cal G}_A(y,x) \right\},
 \ee
where 
\be 
  \label{measQCD}
d\mu_A \ =\ \prod_{x\mu a} dA_\mu^a (x) \prod_f {\rm det}
\| i /\!\!\!\!{\cal D}^E  -  m_f\| 
\exp \left\{ - \frac 1{4g^2} \int F_{\mu\nu}^a   F_{\mu\nu}^a  d^4x \right \}
  \ee
is the standard QCD 
measure and ${\cal G}_A(x,y)$ is the Euclidean Green function
of the $u$ and $d$ quarks in a given gauge-field background. 
Note that, when writing down Eq. (\ref{TrGSGS}), 
we used the fact that 
$J^{\bar u d}$ is not a singlet in flavor [otherwise, the disconnected
 contribution 
$$\propto {\rm Tr} \left\{ \Gamma {\cal G}_A(x,x) \right\}
 {\rm Tr} \left\{ \Gamma {\cal G}_A(y,y) \right\}$$ would appear 
on the right-hand side].
In addition,   the assumption $m_u = m_d$ 
was made [otherwise, we would have two 
different  Green's functions ${\cal G}^u_A(x,y) \neq {\cal G}^d_A(x,y)$].

An important nontrivial relation
 \be
\label{gSg}
\gamma^5 {\cal G}_A(x,y) \gamma^5 \ =\ {\cal G}_A^\dagger (y,x)
 \ee
holds. To understand it, write
the spectral decomposition for  ${\cal G}_A(x,y)$,
  \be
\label{speccont}
{\cal G}_A(x,y) \ =\ \langle \psi (x) \bar \psi (y) \rangle^A \ =\ \sum_{k}
\frac {u_k (x) u_k^\dagger (y) }{m - i \lambda_k},
 \ee 
where $u_k(x)$ are the eigenfunctions of the Dirac operator [cf.\ Eq. (\ref{speclat})]. As we mentioned earlier, for any such  eigenfunction with eigenvalue $\lambda_k$, $\gamma^5 u_k$ is also an eigenfuction with eigenvalue $ - \lambda_k$. This allows us to write:
 \be
 \gamma^5 {\cal G}_A(x,y) \gamma^5  &=&
\sum_{k}
\frac {[\gamma^5 u_k (x)][\gamma^5 u_k (y)]^\dagger }{m - i \lambda_k} 
\ =\  \sum_{p}
\frac {u_p (x) u_p^\dagger (y) }{m + i \lambda_p}
\nonumber\\[0.1cm] 
&=&
\left[ \sum_{p}
\frac {u_p (y) u_p^\dagger (x) }{m - i \lambda_p}  \right]^\dagger =\, 
 {\cal G}_A^\dagger (y,x) ,
 \ee
as annonced. We see that the pseudoscalar correlator
$$\sim \left \langle {\rm Tr} \{ \gamma^5 {\cal G}_A(x,y) 
\gamma^5 {\cal G}_A(y,x) \} \right
\rangle \ =\  \left \langle {\rm Tr} \{ | {\cal G}_A(x,y)|^2 \} \right
\rangle $$
plays a distinguished role --- it represents an absolute
upper bound for any other such correlator. The fastest way to show 
this is
to expand the $4 \times 4$ matrix ${\cal G}_A(x,y)$ over the full basis
 \be
\label{structS}
{\cal G}_A(x,y) = s(x,y) + \gamma^5 p(x,y) + i \gamma^E_\mu v_\mu(x,y) +
\gamma^E_\mu \gamma^5 a_\mu(x,y)  
+\ \frac 12 i \sigma_{\mu\nu} t_{\mu\nu}(x,y)\ .
  \ee
Then
  \be
- \frac 14 C^A_{\gamma^5}(x,y) &=&
\frac 14   {\rm Tr} \{ \gamma^5 {\cal G}_A(x,y) \gamma^5 {\cal G}_A(y,x) \} 
 \ =\ \frac 14   {\rm Tr} \{ |{\cal G}_A(x,y)|^2 \}\nonumber\\[0.1cm]
& =&
|s|^2 + |p|^2 + |v_\mu|^2 + |a_\mu|^2 + 
 \frac 12 |t_{\mu\nu}|^2,
   \ee
but, say,
   \be
   - \frac 14 C^A_{\bf 1}(x,y) &=&
\frac 14   {\rm Tr} \{  {\cal G}_A(x,y)  {\cal G}_A(y,x) \}
=\frac 14   {\rm Tr} \{ {\cal G}_A(x,y) \gamma^5 
{\cal G}_A^\dagger(x,y)  \gamma^5 \} \nonumber\\[0.2cm] 
& =&
|s|^2 + |p|^2 - |v_\mu|^2 - |a_\mu|^2 + \frac 12 |t_{\mu\nu}|^2  .
 \ee
The inequalities 
  \be
  \label{ineqPS}
|C^A_{\gamma^5}(x,y)| \geq |C_\Gamma^A(x,y)|
  \ee
in any given gauge background, the  positivity of the measure 
(\ref{measQCD}) in Eq. (\ref{TrGSGS}), 
and the asymptotics (\ref{corrasmes}) imply that the mass $M_{PS}$ of the 
lightest pseudoscalar meson in the $\bar u d$ channel is less or may be
equal to the masses $M_S,\, M_V,\, M_A,\, M_T$ of the lightest scalar, vector,
axial, and tensor states.\footnote{We 
were a little bit sloppy here. The  inequalities
(\ref{ineqPS}) hold strictly speaking only for 
unrenormalized correlators, not for the renormalized ones. However, 
renormalization only brings about multiplicative    
factors, which do not depend on distance. 
Thus, sending $|x - y|$ to infinity {\it before} 
the limit $\Lambda_{\rm UV} 
\to \infty$ is done, we can ensure that the inequalities (\ref{ineqPS}) 
for renormalized correlators at large distances are  fulfilled.} 

Let us emphasize again that this statement can be justified only in 
the theory
with the positive measure (\ref{measQCD}) (on the other hand,  in the theory 
with nonzero
vacuum angle $\theta \neq 0 $, the measure {\it is} not positive and 
pseudoscalar
states {\it need} not to be the lightest), with 
equal quark masses, and only for those states that are not flavor-singlet.
For flavor-singlet states it need not be true.  Consider, for example,
the theory with just one quark of a large mass. Then the lowest meson states
would be made of gluons and would know nothing about quarks. The 
lowest glueball
 state is believed to be scalar rather than  pseudoscalar.

In Sect. 6, we 
mentioned already the  Vafa--Witten theorem \cite{Vafa} saying that the
vector isotopic symmetry is not broken spontaneously in QCD. Now we are ready
to prove it. Indeed, if such a breaking occured, 
the massless Goldstone scalar
 particles would appear in the spectrum. The inequality $M_{PS} \leq M_S$
implies that a massless {\it pseudoscalar} particle would also exist. But
the theory with $m_u = m_d \neq 0$  (which duly enjoys the exact isovector
 symmetry,
a possible spontaneous breaking of which is under discussion now) 
has no exact axial isotopic symmetry, and there are {\it no reasons} 
for the massless pseudoscalar state to exist. So, it does not exist, hence 
the 
massless scalar does not exist either, and 
the isovector symmetry is not broken.

Many more inequalities of this kind (e.g.\ $M_N \geq M_\pi$ or $M_{\pi^+}
\geq M_{\pi^0}$) can be formulated, but their proof relies on
some extra assumptions and even though the assumptions are very natural, the status
 of these results is a little bit less solid.  We address the reader to
 Ref. \cite{Nussinov} for a nice review.
 
\subsection{Euclidean Dirac spectral density.}

Consider the Euclidean
 Dirac operator $/\!\!\!\!{\cal D}^E$ in a given gauge 
field background $A_\mu(x)$. We assume that the system is placed in a finite
4-volume so that the spectrum of $/\!\!\!\!{\cal D}^E$ is discrete. Let
$\{\lambda_k\}$ be the background-dependent set of eigenvalues of 
$/\!\!\!\!{\cal D}^E$. The {\it spectral density}\,\footnote{The 
notion of   spectral density and the definition (\ref{specdens})
are also widely used in  condensed matter physics and nuclear physics. It
is especially useful if a system is disordered or involves elements of 
disorder like it is the case for  electron spectra in most solids or for 
energy levels in complicated nuclei. It makes sense 
also for ordered systems (such as 
metals). In this case, rather than averaging
over stochastic external
field, one averages over some interval of eigenvalues
$\Delta \lambda$ much larger than the characteristic level spacing, but much less
than the characteristic scale of $\lambda$ on
which $\rho(\lambda)$ is essentially
changed.} is defined as follows:
   \be
\label{specdens}
 \rho(\lambda) \ =\ \left\langle \frac 1V \sum_k \delta \left( \lambda -
\lambda_k[A_\mu(x)] \right)  \right\rangle \ ,
  \ee
where the average is done with the  weight 
function (\ref{measQCD}).
The $\gamma^5$ symmetry of the spectrum  implies that
$\rho(\lambda)$ is an even function of $\lambda$. 

In contrast to  solids or  nuclei, the spectral density (\ref{specdens}) 
is defined in  Euclidean space and seems to have no direct 
physical meaning. There are, however, a set of remarkable identities which
relate the spectral density of
the Euclidean Dirac operator to  physical 
observables. The simplest such identity relates    the spectral density
at ``zero virtuality'' $\lambda = 0$ to the quark
condensate.

 To derive it, 
  set  $x=y$ in the spectral decomposition (\ref{speccont}), integrate it 
over $\frac 1V d^4x$,
and perform the averaging over the
gauge fields with  weight (\ref{measQCD}).
In view of the definitions \p{psiLR}, (\ref{condfg}), (\ref{conddel}), 
and (\ref{specdens}),
using the symmetry $\rho(-\lambda) = \rho(\lambda)$, and assuming the reality 
of $\Sigma$, we obtain
 \be
 \label{Banks0}
\Sigma   \ =\ \left\langle \sum_k \frac 1{m - i\lambda_k} \right \rangle
\ =\ \int_{-\infty}^\infty \frac {\rho(\lambda) d\lambda }{m - i\lambda} 
\ =\ 2m \int_0^\infty  \frac {\rho(\lambda) d\lambda }{\lambda^2 + m^2} .
 \ee
To  better understand this formula, let us  first look 	at 
what happens for free
fermions. As there is no physical dimensionful scale in this case [remember
that $\lambda_k$ in Eq. (\ref{specdens}) are eigenvalues of the {\it massless}
Dirac operator], $\rho(\lambda) = C|\lambda|^3$ on dimensional 
grounds. By 
counting the eigenvalues of the free Dirac operator,
 \be
 \label{specfrefer}
 \lambda(n_\alpha) \ =\ \pm\frac{2\pi}L \sqrt{\sum_\alpha \left(n_\alpha + \frac 12 
\right)^2}
 \ee
in the 4D ball $1/L\ll |\lambda| < \Lambda$ 
[antiperiodic boundary conditions for the fermions in all four directions are 
chosen, $n_\alpha$ are integer, and each level (\ref{specfrefer}) involves an 
extra $2N_c$-fold
 degeneracy],   it is not
difficult to determine $C = N_c/(4\pi^2)$. Thus,
  \be
\label{densfree}
  \rho^{\rm free}(\lambda)\ =\ \frac {N_c}{4\pi^2} |\lambda|^3 .
  \ee
In  the interacting theory, the spectral density behaves as
$  \rho(\lambda)\ \propto \ |\lambda|^3$ for $|\lambda|$  much 
greater than the characteristic 
hadron scale $\mu_{\rm hadr}$, so that 
interaction is weak due to asymptotic freedom. To be more precise, the power $|\lambda|^3$ is
multiplied (in the leading logarithmic order) by the anomalous
 dimension factor 
  \be
 \label{andimlam}
\sim \left[ \frac {\alpha_s(|\lambda|)}{\alpha_s(\mu)}  \right]^{\gamma/b_0} ,
 \ee
where $\mu \sim \mu_{\rm hadr}$ is the normalization point and
$b_0 = 11 - 2N_f/3$ was defined in Eq. \p{b0}.
According to the calculations in Ref. \cite{Neveu},
 $\gamma = 16$. 

We see that the integral in Eq. (\ref{Banks0})    
diverges quadratically in the
ultraviolet. The same result can be obtained directly by calculating the 
fermion bubble graph in the momentum representation
 \be
\label{condUV}
\langle \psi(0) \bar \psi (0) \rangle \ = \ \int \frac{d^4p_E}{(2\pi)^4}
 \ {\rm Tr}
\left\{ \frac{/\!\!\!{p_E} + m}
{p_E^2 + m^2} \right\} \ \propto \ m \Lambda_{UV}^2 .
 \ee
Thus, strictly speaking, the formula (\ref{Banks0}) does not make much sense
as it stands. Note, however, that even though the (purely perturbative) 
contribution (\ref{condUV}) diverges in 
the ultraviolet, it vanishes in the 
chiral limit $m \to 0$. The whole point is that in QCD the integral
(\ref{Banks0}) acquires an additional {\it nonperturbative} contribution
coming from the region of small $\lambda$ which survives in the 
``continuum chiral  thermodynamic limit''
 ({\it first} $V \to \infty$, 
{\it then} $m \to 0$, and only {\it then} the ultraviolet cutoff is lifted
$\Lambda_{UV} \to \infty$). The fact that chiral symmetry is broken 
spontaneously {\it means} that the vacuum expectation value 
$ \langle \psi(0) \bar \psi (0) \rangle $ is nonzero in this 
particular limit.

Obviously, the necessary condition for the condensate
 to develop is $\rho(0) \neq 0$. Neglecting all
terms which vanish in the continuum  chiral  
thermodynamic limit
  defined
above, we  finally obtain the famous {\it Banks--Casher 
relation} \cite{Banks}
 \be
\label{Banks}
 \langle \psi(0) \bar \psi (0) \rangle_{\rm vac}\ \equiv \ \Sigma \ =\ \pi 
\rho(0).
  \ee
Note that the result does not depend on flavor, which tells us again that the
flavor vector symmetry is not broken.

\vspace{1mm}

Not only $\rho(0)$, but also the form of $\rho(\lambda)$ at small $\lambda 
\ll \mu_{\rm hadr}$ can be determined \cite{Stern}. 
Consider the theory with $N_f \geq 2$
light quarks of common mass $m$. Let us study  the integrated 
correlator 
 \be
 \int d^4x \, \langle S^a(x) S^b(0) \rangle \ =\ \frac 1V \int d^4x d^4y
  \, \langle S^a(x) S^b(y) \rangle ,
  \ee
where $S^a(x) = \bar \psi(x) t^a \psi(x)$ and $t^a$ is the generator of the
SU$(N_f)$ flavor group. Fix a particular gluon background and define
  \be
\left.  C^{ab} \right|_A \ =\ - \frac 1V \int d^4x d^4y \, {\rm Tr}
\left\{ t^a  {\cal G}_A(x,y) t^b  {\cal G}_A(y,x) \right\}.
\label{CabA}
\ee
Substitute here the spectral decomposition (\ref{speccont}) for 
$ {\cal G}_A(x,y)$, do the integration,  and perform  averaging over 
the gluon
fields trading the sum over eigenvalues for the integral 
over the spectral
density (\ref{specdens}). We obtain
  \be
 \label{Cabint}
C^{ab} &=& - \frac{\delta^{ab}}{2V} \left\langle \sum_k 
\frac 1{(m - i\lambda_k)^2} \right \rangle\   = \  - \frac {\delta^{ab}}{2}
\int_{-\infty}^\infty \frac {\rho(\lambda) d\lambda }{(m - i\lambda)^2}
 \nonumber\\[0.2cm]  
&=&-{\delta^{ab}} 
\int_{0}^\infty \frac {\rho(\lambda)(m^2 - \lambda^2)  }{(m^2 + \lambda^2)^2}
 d\lambda  ,
 \ee
where the property $\rho(-\lambda) = \rho(\lambda)$ was used.

\begin{figure}
\bc
 \includegraphics[width=.35\textwidth]{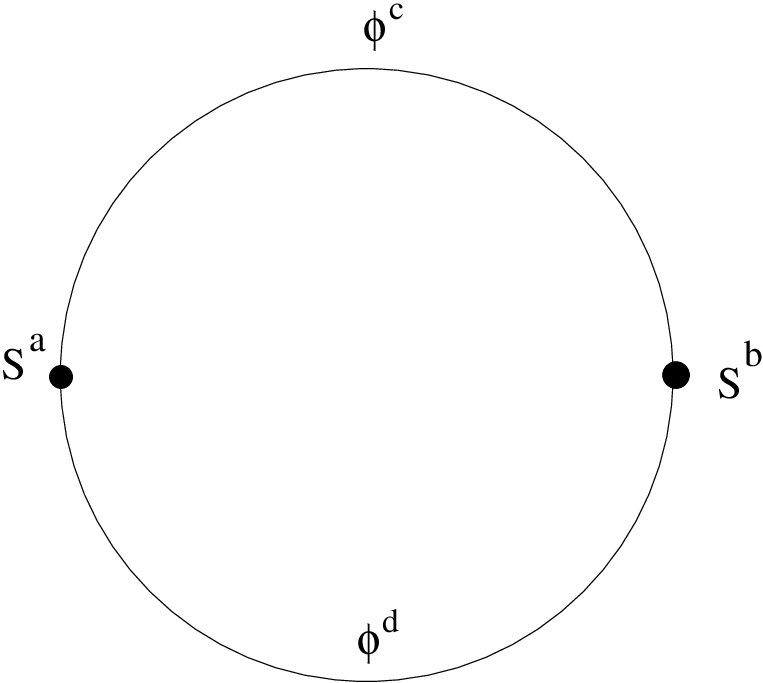}
\ec   
\caption{Pseudogoldstone loop in  scalar correlator.}
\label{Sternfig}
\end{figure}

On the other hand, the same correlator can be saturated by physical states,  
among which the lightest pseudo-Goldstone
 states play a distinguished role. 
Consider the 1-loop graph in Fig. \ref{Sternfig} describing the contribution
 of 
the 2-goldstone
 intermediate state 
$\sim \int \langle 0|S^a |\phi^c \phi^d \rangle 
\langle \phi^c \phi^d | S^b | 0 \rangle $ (obviously, the one-particle
state does not contribute, because the pseudo-Goldstone
mesons are pseudoscalars
while $S^a(x)$ is scalar). To calculate it, we need to know the vertex
$\langle 0|S^a |\phi^c \phi^d \rangle$, which can be determined 
via the 
generating functional
 of QCD involving  scalar sources $u^a$ coupled to the
current $S^a$. 
Adding the source term $u^a S^a$ to the Lagrangian amounts 
to adding $u^a t^a$
to the quark mass matrix ${\cal M}$.  The latter also enters  the mass term
(\ref{Lchirmass}) in the effective Lagrangian.
  Expanding $U$ up to  second
order in $\phi^a$ and varying ${\cal L}_{\rm eff}^{(m)}$ with respect to 
$u^a$, we derive 
 \be
 \label{Sfifi}
\langle 0|S^a |\phi^c \phi^d  \rangle \ =\  -\frac {\Sigma}{F_\pi^2} 
d^{abc},
 \ee
 with $d^{abc}$ being a symmetric group tensor:
\be
d^{abc} \ =\ 2{\rm Tr} \{ t^a(t^b t^c + t^c t^b) \}.
\ee
The vertex is nonzero only for three or more flavors. 

Now we can calculate the graph in Fig. \ref{Sternfig}. Actually, 
we cannot because
the integral diverges logarithmically in the
ultraviolet, but anyway  the
effective theory is not valid at high momenta (technically, the divergence is 
absorbed into local counterterms
 of higher order in $p^{\rm char} $ and $m$).
Only the infrared-sensitive part of the integral is relevant. A simple 
calculation using the identity 
$$
d^{abc} d^{abd}  \ =\ \frac {N_f^2-4}{N_f} \delta^{cd} 
$$
 gives
 \be
 \label{Cabinfr}
\left( C^{ab} \right)^{\rm infrared} \ =\ - \frac{N_f^2 - 4}{32 \pi^2 N_f}
\left( \frac \Sigma {F_\pi^2} \right )^2  \delta^{ab} \ln \frac
{M_\phi^2}{\mu_{\rm hadr}^2} .
  \ee
Now compare it  with Eq. (\ref{Cabint}). Note first of all that the constant 
part $\rho(0)$ does not contribute here,
$$ \int_0^\infty \frac {m^2 - \lambda^2}{(m^2 + \lambda^2)^2} d\lambda \ =\ 0 
\ . $$
Thus, only the difference $\rho(\lambda) - \rho(0)$ is relevant. It is easy
to see that, in order to reproduce the singularity $\sim \ln M_\phi^2 \sim
\ln m$, we should have  $\rho(\lambda) - \rho(0) = C|\lambda|$ at small
 $|\lambda|$. Substituting it in Eq. (\ref{Cabint}) 
and comparing the coefficient of $\ln m$ with 
the coefficient of $\ln M_\phi^2$ in Eq. (\ref{Cabinfr}), we
finally obtain$\,$\cite{Stern}
 \be
\label{SmStern}
 \rho(\lambda) \ =\ \frac \Sigma \pi + \ \frac{N_f^2 - 4}{32 \pi^2 N_f}
\left( \frac \Sigma {F_\pi^2} \right )^2 |\lambda| \ + o(\lambda).
 \ee
The behavior is smooth in the theory with two light flavors, but for $N_f \geq 3$, $\rho(\lambda)$  has a nonanalytic ``dip'' at
$\lambda = 0$. 
Physically, it is rather natural that the greater the number of flavors, the 
stronger the suppression of 
$\rho(0)$  is. The determinant factor in the measure (\ref{measQCD})
punishes small eigenvalues, and the larger  $N_f$ is, the more important 
this factor becomes. 
By ``analytic continuation'' of this argument, one should expect
a nonanalytic bump rather than a nonanalytic dip at $\lambda = 0$ in the 
case $N_f = 1$. Indeed, Eq. (\ref{SmStern}) displays such a bump. 
One should not forget, of course, that the whole derivation was based on the 
effective chiral Lagrangian
 approach
and does not directly apply to the case $N_f = 1$.
Some additional, more elaborate reasoning based on the random matrix model shows, however, 
that a bump at $N_f = 1$, as predicted by Eq. (\ref{SmStern}), 
is there \cite{Toublan}. 
The existence of the bump was also confirmed by a numerical calculation in the instanton liquid model
 \cite{Verb} (see the plots in Fig. 1 there). 

\subsection{Infrared face of anomaly.}

As was explained in Sect. 3, the symmetry (\ref{U1chir}) of the classical
theory involving massless fermions  is broken down by quantum effects. The
$U_A(1)$ breaking is introduced by an ultraviolet regularization [so that the
measure in the functional integral is not $U_A(1)$ invariant] and the 
effects due to this breaking do not go away in the limit 
$\Lambda_{UV} \to \infty$. This is 
the ultraviolet face of the anomaly.
 The latter  also has, however, a 
different,  
infrared face: one can understand its origin exclusively in terms of the low
energy dynamics of the theory \cite{Dolgov}. We have already seen how the anomaly is related
to the presence of fermion zero modes
 in an Euclidean topologically nontrivial
gauge background. Let us  now find out what happens in  Minkowski  space.

 \begin{figure}
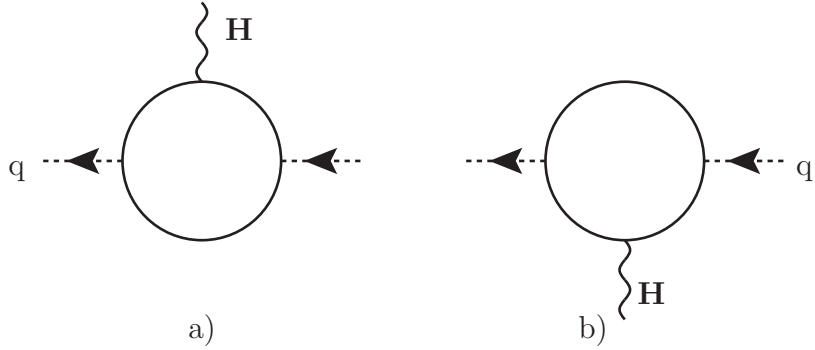

   \begin{center}
      \begin{axopicture}(210,110)
         \SetScale{2}

           \Text(30,-2)[c]{a)}

          \DashArrowLine(15,30)(0,30){1}
           \BCirc(30,30){15}
          \DashArrowLine(60,30)(45,30){1}
           \Photon(30,45)(30,60){1}{2}

          \Text(-5,28)[c]{q}
          \Text(37,55)[c]{{\vec H}}
          
          \Text(104,-2)[c]{b)}

          \DashArrowLine(95,30)(80,30){1}
             \BCirc(110,30){15}
          \DashArrowLine(140,30)(125,30){1}
            \Photon(110,15)(110,0){1}{2}

          \Text(144,28)[c]{q}
          \Text(115,5)[c]{{\vec H}}

     \end{axopicture} 
   \end{center}
\caption{Anomalous triangle in diangle kinematics.}
\label{treugfig}
\end{figure}

Let us discuss the massless QED first. Consider the correlator
 \be
 \label{VAxH}
T_{\mu\nu}^{ H} (q) \ = \  i \int \langle T\{j_{\mu 5}(x) j_\nu(0) \} 
\rangle_{\vecind H} \ e^{iq\cdot x} d^4x
 , \ee
where $j_\nu = \bar \psi \gamma_\nu \psi,\ j_{\mu 5} = \bar \psi \gamma_\mu
\gamma^5 \psi$, and the averaging is performed in the presence of an external 
homogeneous magnetic field ${\vec H}$. The correlator can be calculated 
perturbatively as a series in $e^2$ and    ${\vec H}$.  The vacuum correlator $ \langle T\{j_{\mu 5}(x) j_\nu(0) \} 
\rangle_{\rm vac} $ vanishes and the expansion in ${\vec H}$ starts with the
linear term. The latter is given by the graphs in Fig. \ref{treugfig}, where the 
wavy line signals the modification of the electron propagator due to the external 
field to leading order, 
 \be
\Delta G(p) \ =\  e\varepsilon_{ijk} H_k \frac {\gamma_i (/\!\!\!{p} - m) 
 \gamma_j}{2(p^2 - m^2)^2} 
\ee
In the massless limit it coincides with the second term in Eq. (\ref{GAp}) (In this subsection, devoted to infrared perturbative aspect of the anomaly, we do not include the charge $e$ in the definition of the field). Actually, the graphs in Fig. \ref{treugfig} describe 
the 3-point correlator
 \be
 \label{T3point}
T_{\mu\nu\alpha} (q, k) \ = \ i \int \langle T\{j_{\mu 5}(x) j_\nu(0) 
j_\alpha(y)\} 
\rangle \ e^{iq\cdot x - ik\cdot y} d^4x d^4y
\ee
in a special kinematics: $\alpha$ is spacelike and  $k = (0, {\vec k} \to {\vec 0})$. This 
kinematics is somewhat
 simpler and  more instructive theoretically than the standard symmetric kinematics
$k^2 = (q-k)^2 = 0$ used in Ref. \cite{Dolgov} (which on the other hand is better adapted to describe the
 phenomenology of the decay $\pi^0 \to \gamma \gamma$).  To make things
 still simpler, we assume that the vectors ${\vec q}$ and ${\vec H}$ are parallel
 and direct them  along
the 3rd axis. Calculating the integral with, say, 
the Pauli--Villars regularization
 method,\footnote{Calculational details 
will be given later in the solution of  {\bf Problem 7} for
the two-dimensional case, which is somewhat simpler.}
 we obtain
 \be
 \label{TmnH}
T_{\mu\nu}^{ H} (q) \ = \  \frac {e{ H}}{2\pi^2} \frac {q_\mu 
\tilde{\varepsilon}_{\nu\alpha} q^\alpha}{q^2}
 , \ee
with $H = |{\vec H}|$ and $\tilde{\epsilon}_{\nu\alpha}$ living in the 
two-dimensional (03)-space so that $\tilde{\varepsilon}_{03} = - \tilde{\varepsilon}_{30} = -1$, and
$\tilde{\varepsilon}_{\perp \alpha} = 0$.

The amplitude (\ref{TmnH}) satisfies the property 
$q^\nu T_{\mu\nu}^{ H} \ =\ 0$,  which reflects the conservation of the vector
current. On the other hand,
 \be 
\label{divTmn}
q^\mu T_{\mu\nu}^{ H} \ = \ \frac{eH}{2\pi^2} \tilde \varepsilon_{\nu\alpha} 
q^\alpha \ \neq \ 0,
 \ee
and this is a manifestation of the anomaly (\ref{abanom}). In fact, 
Eq. (\ref{divTmn})
means that the average of the operator $\partial^\mu j_{\mu 5}$ in the presence of a  
magnetic {\it and} an electric field $E_i = \partial_0 A_i - \partial_i A_0$ is 
equal to\footnote{
To derive the operator relation \p{divEH} out of \p{divTmn}, it is more convenient again to treat a little more simple $2D$ case first and then generalize to four dimensions. We will do so on the next page.}
\be
\lb{divEH} 
\langle \pd^\mu j_{\mu 5} \rangle \ =\ -\frac {e^2}{2\pi^2} \vecg{E} \cdot \vecg{H} ,
 \ee
which is the QED version of Eq. \p{chiranom}.
The important observation is that the amplitude (\ref{TmnH}) is singular
at $q^2 = 0$, and this singularity can only be explained  by the presence
of  {\it massless particles} in the spectrum. It is very instructive to see
what happens if electrons are endowed with a small mass, which 
explicitly  breaks   the $U_A(1)$ invariance and also smears out the
singularity in Eq. (\ref{TmnH}). The direct calculation of Im 
$  T_{\mu\nu}^{ H, m}$ by the graphs in Fig. \ref{treugfig} with nonzero
mass gives\footnote{The appearance of the factor $m^2$ in the numerator of the middle term in Eq. \p{ImTdel} can be heuristically understood as follows. This numerator represents a product  of  the amplitude of production of $q\bar q$ pair by the axial source and the amplitude of transition of this pair into two photons. 

If the quarks produced by the axial source were strictly massless, their total spin would be 1. But then they could not go into two photons: $\gamma\gamma$ system cannot have total anglular momentum 1 \cite{BLP}.} 
  \be
 \label{ImTdel}
{\rm Im } \ T_{\mu\nu}^{H, m} \ =\ -\frac{eH}{\pi} 
\frac {m^2 \theta(q^2 - 4m^2)} {\sqrt{q^2 (q^2 - 4m^2)}}
 \frac {q_\mu  \tilde{\varepsilon}_{\nu\alpha} q^\alpha} {q^2} \ 
\stackrel{m \to 0}{\longrightarrow} \ 
 -\frac{eH}{2\pi} \delta(q^2) q_\mu  \tilde{\varepsilon}_{\nu\alpha} q^\alpha.
 \nonumber\\
 \ee
This means that anomalous
 nonconservation of the axial charge in massless QED is
associated with the creation of massless $e^+e^-$  pairs of zero 
energy 
in the presence of electric and magnetic fields with 
${\vec E} \cdot {\vec H} \neq 0$. These
pairs carry  nonzero axial charge. If ${\vec E}$ and ${\vec H}$ are 
constant and homogeneous, the pairs are created all the time and everywhere. 
If the fields die out fast enough at spatial infinity and also in the limits
$t \to \pm \infty$, 
 the number of  created pairs  is finite and coincides with the change of the
 axial charge which, according to Eq. (\ref{divEH}), is equal to \cite{Ambj}: 
 \be
 \label{delQ5}
\Delta Q_5 =  -\frac {e^2}{2\pi^2} \int d^3x dt \ 
{\vec E} \cdot {\vec H} .
 \ee

\vspace{.4cm}
{\bf Problem 7}. Discuss a diagrammatic interpretation 
of the axial anomaly
 (\ref{anomd2}) in the 2-dimensional QED
({\it Schwinger
 model}).

\vspace{.1cm}
{\bf Solution}. The right side of Eq. (\ref{anomd2}) is linear in field, which
means that the anomaly shows up in the vacuum 2-point correlator
  \be
 \label{TmnSM}
T_{\mu\nu}^{\rm 2D} (q) \ = \ i \int \langle T\{j_{\mu 5}(x) j_\nu(0) \} 
\rangle \ e^{iq\cdot x} d^2x  .
  \ee
Thus, instead of the graphs in Fig. \ref{treugfig}, we have just one 
anomalous diangle.
 In spite of the fact 
that the corresponding Feynman integral converges in the ultraviolet
 (it diverges logarithmically by power counting, but the leading term 
 $\propto {\rm Tr} \{\gamma_\mu\gamma^5 /\!\!\!{p} \gamma_\nu /\!\!\!{p} \}$
 vanishes after integration), it still has to be regularized (otherwise, the property
$q^\nu T_{\mu\nu} = 0$ required by gauge invariance would not hold). We will choose  Pauli--Villars
 regularization.  
Subtracting  the heavy fermion loop, we obtain for small $m^2 \ll |q^2|$ (not forgetting the factor $-1$ due to the fermion loop) the expression
   \be
    \label{qTmnSM}
    T_{\mu\nu}(q) \ =\  -i \ \int \frac {d^2p}{(2\pi)^2}
    \left[ {\rm Tr} \left\{  \gamma_\mu\gamma^5 
    \frac {i/\!\!\!{p} } {p^2 } \gamma_\nu  \frac {i(/\!\!\!{p} + /\!\!\!{q}) }    
{(p+q)^2 } \right\}\right. 
\nonumber \\ 
 +\  \left. \frac {M^2}{(p^2-M^2)^2} {\rm Tr} 
    \{ \gamma_\mu\gamma^5  \gamma_\nu \} \right] .
  \ee 
The calculation gives:
 \be
\lb{TmnSMan}
 T_{\mu\nu}(q) \ =  \frac 1{2\pi} \left[ \frac {(\varepsilon_{\mu\alpha} q_\nu
+ \varepsilon_{\nu\alpha} q_\mu) q^\alpha}{q^2}  - \varepsilon_{\mu\nu} \right]
  \ =\  \frac {q_\mu \varepsilon_{\nu\alpha} q^\alpha}{\pi q^2},
 \ee
  where the we used the convention $\varepsilon^{01} = - \varepsilon_{01} = 1$, the property $\eta_{\beta\alpha} \varepsilon_{\mu\nu} + \eta_{\beta\mu} \varepsilon_{\nu\alpha} +
  \eta_{\beta\nu} \varepsilon_{\alpha\mu} \ =0$ and chose the $2D$ gamma matrices in the form \p{2Dgamma}, giving\footnote{Note that $\gamma^\alpha \gamma_\beta \gamma_\alpha = 0$ in two dimensions.}

\be
{\rm Tr}  \{ \gamma_\mu \gamma_\nu \gamma^5 \}  \ &=&\  2 \epsilon_{\mu\nu}, \nn
{\rm Tr}  \{ \gamma_\mu \gamma_\nu \gamma_\alpha \gamma_\beta \gamma^5 \}  \ &=&\  \epsilon_{\mu\nu}\, \eta_{\alpha\beta}+ 
\epsilon_{\alpha\beta}\, \eta_{\mu\nu} +  \epsilon_{\alpha\nu}\, \eta_{\mu\beta} +  \epsilon_{\mu\beta}\, \eta_{\alpha\nu} . 
 \ee 
 
 \begin{figure}
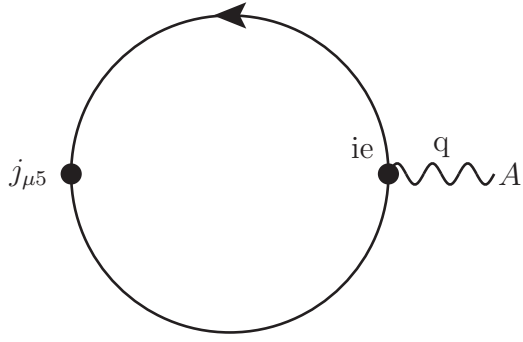

   \begin{center}
      \begin{axopicture}(210,110)
         \SetScale{2}
\Arc[arrow](30,30)(30,0,180)
\Arc (30,30)(30,180,360)
\Vertex(0,30) {2}
\Photon[arrow](60,30)(80,30) {2} {3}
\Text(70,35){q}
\Vertex(60,30) {2}
\Text(83,30) {$A$}
\Text(55,35) {ie}
\Text(-8,30){$j_{\mu 5}$}
          \end{axopicture} 
   \end{center}
\caption{$\langle j_{\mu 5} \rangle_E $ in Schwinger model.}
\label{j5A}
\end{figure}

Consider now the average value $\langle j_{\mu 5} \rangle_E$ in the presense of the background electric field $E = - \pd_1 A_0$.  It is described by the diagram in Fig. \ref{j5A}, which gives:
\be
\langle j_{\mu 5} \rangle_E \ =\ -i T_{\mu\nu}(q) (ie) A^\nu(q) \ =\ \frac {eq_\mu \varepsilon_{\nu\alpha}q^\alpha A^\nu(q)}{\pi q^2}.
\ee 
Multiplying this by $q^\mu$ and going over into  coordinate space, we derive:
\be
\lb{dmjm5}
\langle \pd^\mu j_{\mu 5} \rangle_E  \ =\ \frac e\pi \varepsilon^{\nu\alpha} \pd_\alpha A_\nu \ =\ - \frac E\pi = -\frac e{2\pi} \varepsilon^{\alpha\nu} F_{\alpha \nu}.
\ee
 Absorbing the charge $e$ in the definition of the field, we arrive at \p{anomd2}.

\vspace{1mm}

Going back to four dimensions, one can observe that
the result (\ref{TmnH})
  coincides with   Eq. (\ref{TmnSMan}) 
  up to the change 
${\varepsilon}_{\nu\alpha} \to   \tilde {\varepsilon}_{\nu\alpha}$ and with an extra
factor $eH/2\pi$.  The relations (\ref{divTmn})--(\ref{ImTdel}) also have their exact two-dimensional counterparts.  

In particular, the anomaly relation \p{divEH}
can be derived out of \p{divTmn} in the same way as we derived \p{anomd2} out of \p{TmnSMan}.  To this end, it is sufficient to consider the kinematics when $\vecg{E} \| \vecg{H}$ and both fields are directed along the 3rd axis. Then on the place of \p{dmjm5}, we obtain \p{divEH}.

 The physical interpretations  in $4d$ and $2d$  are practically identical: an external electromagnetic field with nonzero $\int dt d^3 x\, \vecg{E} \cdot \vecg{H}$ in $4d$ or  an electric field with nonzero $\int dt dx \, E$ in $2d$ 
bring about the change of axial charge associated with the creation of soft massless fermion-antifermion  pairs.

\vspace{1mm}

As a side remark, note that in two dimensions vector and axial currents are  
related as $j_{\mu 5} = -\varepsilon_{\mu\nu} j^\nu$, and the singularity
of the correlator (\ref{TmnSM}) means also the singularity of the vector 
polarization operator $\Pi_{\mu\nu} \propto 
(\eta_{\mu\nu} - q_\mu q_\nu /q^2)$, i.e.\ nonvanishing $\Pi(0)$. As a result, the Schwinger photon 
acquires the  mass, 
 \be
\label{SMmass}
\mu^2 = g^2/\pi . 
 \ee
Actually, the only physical states in the massless Schwinger model
 are {\it free  bosons} with mass (\ref{SMmass}). And that means that the fermion fields 
entering the Lagrangian of the model are confined!

\subsection{Chiral symmetry breaking and confinement.}

Look
 again at Eq. (\ref{chiranom}). The axial current entering
the left-hand side 
is an external current in a sense that no dynamical field is
coupled directly to $j^{\mu5}$. But the fields entering on the right-hand
 side are
dynamical  gluon fields present in the QCD Lagrangian. 

In the chiral (left-right asymmetric) gauge theories like the standard 
electroweak model, both vector and axial currents are coupled directly to
the physical gauge fields. Anomalous
 divergence of 
such current would mean explicit
breaking of the gauge invariance, which is not nice. Therefore, in chiral 
theories one should always take care 
that such purely {\it internal anomalies}
 would 
cancel out at the end of the day. In the Standard Model,
 they do.

Let us discuss, however,  {\it external} anomalies
 in QCD which are not
related to  breaking  gauge symmetry but only mean that certain correlators involving 
external currents are singular.

As a simplest nontrivial example, consider the theory with two
 massless flavors 
and look at the correlator
  \be
 \label{KabH}
K_{\mu\nu}^{ab, {\cal H}} (q) \ = \ i \int \langle T\{j_{\mu 5}^a(x) 
j_\nu^b(0) \} \rangle_{\cal H} \ e^{iq\cdot x} d^4x,
 \ee
where $a,b$ are flavor indices and ${\cal H}$ is the external 
flavor-singlet ``magnetic field''.\footnote{The 
quotation marks distinguish ${\cal H}$, which is an isosinglet that couples to $B/3$, $B$ being the 
baryon charge, from the physical magnetic
field, which has the matrix structure diag$(2/3, -1/3)$ and is a mixture of 
isotriplet and isosinglet. But we are not interested in dynamics of 
electromagnetic
or weak interactions here. In QCD proper, all color-singlet currents are external.
 ${\cal H}$
is just a source of such vector flavor-singlet current.}
The correlator (\ref{KabH}) is nothing but a three-point vacuum expectation
value (\ref{T3point}) in the  kinematics where one of 
the external momenta 
associated with the vector current is set to zero.

The one-loop calculation of the corresponding graph displays  a
 singularity,
 \be
\label{Kabres}
 K_{\mu\nu}^{ab, {\cal H}} (q) \ = \ 
  -\frac {{\cal  H}}{2\pi^2} \frac {q_\mu 
\tilde{\epsilon}_{\nu\alpha} q^\alpha}{q^2} \times N_c \times \frac 12 
\delta^{ab} ,
 \ee
where the last two factors come from the color and flavor trace. The imaginary part of this amplitude is also singular,
$\sim \delta(q^2)$, which can be related to the masslessness of 
quarks. 
However, the quarks (in contrast to electrons in QED) do not exist as physical
particles  due to confinement and one can ask where does the singularity
 in $ K_{\mu\nu}^{ab,{\cal H}} (q) \propto 1/ q^2$ 
come from?
This is a good question, and the answer is   still better: the singularity
$\sim 1/q^2$ comes from the propagator 
of a massless Goldstone boson,
 which
appears due to the spontaneous 
chiral symmetry breaking  and which is directly
coupled to the axial current $j_{\mu 5}$ (see the middle graph in 
Fig. \ref{Hooftfig}).

\begin{figure}
\bc
 \includegraphics[width=.8\textwidth]{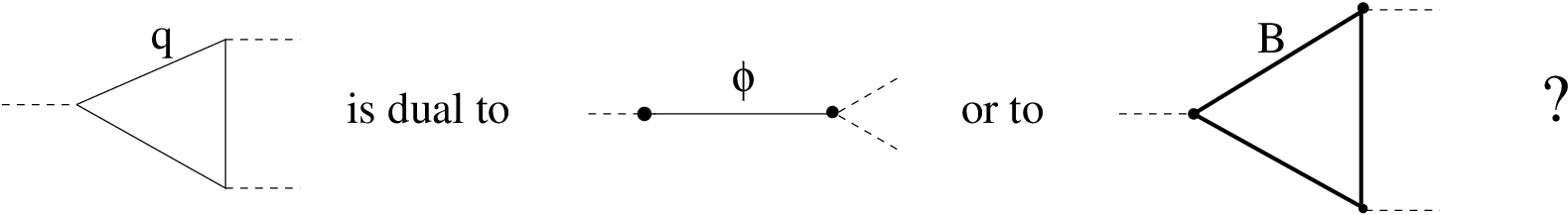}
\ec   
\caption{Saturating the external anomaly.}
\label{Hooftfig}
\end{figure}

Let us ask now:  can one reproduce the singularity in  
Eq. (\ref{Kabres}) {\it without } Goldstone bosons
 and without spontaneous
chiral symmetry breaking, but in some other way?

As far as the theory with two
 light quarks is concerned, the answer is positive:
the singularity of the correlator above can be reproduced,
in principle, if 
{\it massless baryons} are present. Proton and neutron represent, like quarks,
a flavor SU$(2)$ doublet. There are $N_c = 3$ quark doublets and only one
baryon doublet $|P\rangle = |uud\rangle$ and   $|N\rangle = |udd\rangle$. 
The absence of the overall
$N_c$ factor is compensated, however, by the fact that the baryon charge
of the nucleon is 3 times larger than that of  the quark, and the vertex involving
the ``magnetic field"  ${\cal H}$ is 3 times larger for 
 baryons.

Thus,  this purely algebraic   {\it anomaly matching}
 argument due to
 't Hooft  \cite{matching} does not rule out a dynamical
scenario where the physical spectrum in the theory with just two massless quark
flavors would not involve massless pions but, instead, the 
massless proton and neutron. 
It is rather remarkable that, in the theories with $N_f \geq 3$, the scenario
with massless baryons {\it is} ruled out. Suppose that, instead of the octet
of massless Goldstone fields, we have an octet of massless baryons. The 
contribution of the corresponding triangle graph in (\ref{KabH}) would have
the same structure as in Eq. (\ref{Kabres}), but with the factor
 \be
{\rm Tr} \ \{T^a T^b \} \ =\ C_8 \delta^{ab}
 \ee
instead of $\delta^{ab}/2$, where $T^a$ are the flavor generators in the octet
representation. To find the {\it Dynkin index} $C_8$
 of the octet representation, it 
is sufficient to assume that $a=b=1$ and decompose the octet with respect
to the SU$(2)$ flavor subgroup: ${\bf 8} = {\bf 3} + {\bf 2} + {\bf 2} + 
{\bf 1}$. The contribution of each doublet to $C_8$ is  $1/2$ and the 
contribution of the 
triplet is $2$. Adding it together, we obtain $C_8 = 3$, 
which is not the same as $1/2$, and the required result (\ref{Kabres}) is {\it not} reproduced. Also, 
a massless decuplet and all other possible color-singlet baryon 
representations would give the coefficient 
in front of  the singularity much larger than
that in Eq. (\ref{Kabres}), and the anomaly
 matching condition would not be 
fulfilled. Therefore, massless baryons do not exist.\footnote{Well, one could, in  principle,  think of saturating
the anomaly with {\it several} massless baryon multiplets with positive and negative
baryon charges. This possibility is so unaesthetic, however, that it can be 
rejected simply by that reason. In addition, it is ruled out by a careful algebraic analysis \cite{kitaec}.}}

We have arrived at a remarkable 
result. In QCD with three massless quarks, the 
assumption of confinement, allowing the existence of only colorless states in
the physical spectrum, {\it and} the anomaly
 matching condition lead to the
conclusion that massless Goldstone states {\it should}  appear and chiral symmetry {\it must} 
be spontaneously broken. If the latter is not true, the only
possibility to saturate the anomaly
 is to assume that massless quarks still
exist as physical states in the spectrum and there is no confinement!

In the real world, confinement and spontaneous 
chiral symmetry breaking 
in the limit of massless $u$, $d$,
and $s$ quarks are experimental facts. 
Whether or not these 
phenomena 
take place  in hypothetic theories with $N_f \geq 4$  is an open
question. It is quite possible that starting, say, from $N_f \stackrel{?}= 6 $, the small
eigenvalues in the Dirac operator spectrum are punished 
 by the 
determinant   factor so strongly that $\rho(0) =0$ and, in view of the 
Banks--Casher 
relation (\ref{Banks}), the quark
 condensate vanishes  and the symmetry is not
 broken. The   anomaly matching argument tells then that there is no 
confinement in this case.\footnote{We know, 
of course, that if the number of the quark flavors is {\it
very} high $N_f > 16$, the asymptotic freedom
 is lost and we cannot expect
confinement. In addition, we are almost sure that quarks and gluons are not confined
at $N_f = 16$ or $N_f = 15$, in which case the theory is asymptotically free, 
but has the infrared fixed point  at a small value
of $\alpha_s$,
and the coupling constant never grows large.
The argument about suppression of the small Dirac eigenvalues and the results of 
some numerical simulations indicate 
 that confinement might be lost at a  smaller value of $N_f$, not just $N_f = 15$.}  

We have called this result ---
that the chiral
 symmetry breaking and confinement go 
together ---  remarkable.
It is also somewhat mysterious. Even though we do not
understand well dynamical reasons for confinement to occur, we still expect
that {\it the same} mechanism which works for the theory with $N_f=3$ works
also for the theory with $N_f = 2$. But  't Hooft's argument
 works only
for $N_f \geq 3$...

    \section*{Akcnowledgements}
    
    I am indebted to R. Narayanan and M. Zubkov for illuminating discussions.

\end{document}